\documentclass[aps,pra,reprint,nofootinbib]{revtex4-1}

\pdfoutput=1
\usepackage{graphicx}
\usepackage{epstopdf}
\usepackage{amssymb}
\usepackage{mathrsfs}
\usepackage{amsmath,mathtools,amsfonts,bm,graphicx}
\usepackage{upgreek}
\usepackage{color}
\usepackage{etoolbox}

\newcommand{\refb}[1]{(\ref{#1})}
\newcommand{\fig}[1]{Fig.~\ref{#1}}
\newcommand{\Fig}[1]{Figure~\ref{#1}}
\newcommand{\eq}[1]{Eq.~(\ref{#1})}
\newcommand{\eqr}[2]{Eqs.~(\ref{#1})-(\ref{#2})}

\newcommand{\eqs}[2]{Eqs.~(\ref{#1}) and (\ref{#2})}

\newcommand{\Eq}[1]{Equation~(\ref{#1})}
\newcommand{\App}[1]{Appendix~\ref{#1}}
\renewcommand{\sec}[1]{Sec.~\ref{#1}}
\newcommand{\Sec}[1]{Section~\ref{#1}}

\definecolor{darkred}{rgb}{0.6,0,0}

\newcommand{\phiv}{\phi_v}
\newcommand{\w}{\tilde w}
\renewcommand{\l}{\tilde l}
\newcommand{\Vij}{V_{ijs_is_j}}
\newcommand{\Vijt}{V_{ijs_i(t)s_j(t)}}
\newcommand{\Vkl}{V_{ijkl}}
\newcommand{\Vgamma}{V_\gamma}
\newcommand{\Rv}{R_{void}}

\newcommand{\Vol}{\mathcal{V}}

\newcommand{\barm}{\bar{m}}
\newcommand{\barmv}{\phiv \Vol}

\newcommand{\CDE}{\mathcal{C}}
\newcommand{\DEmax}{\CDE k_B T}

\newcommand{\png}{png}
\begin{document}
\title{Local random configuration-tree theory for string repetition and facilitated dynamics of glass}
\author{Chi-Hang Lam}
\email[Email: ]{C.H.Lam@polyu.edu.hk}
\affiliation{Department of Applied Physics, Hong Kong Polytechnic University, Hong Kong, China}
\date{\today}
\begin{abstract}
We derive a microscopic theory of glassy dynamics based on the transport of voids by micro-string motions, each of which involves particles arranged in a line hopping simultaneously displacing one another. Disorder is modeled by a random energy landscape quenched in the configuration space of distinguishable particles, but transient in the physical space as expected for glassy fluids. We study the evolution of local regions with $m$ coupled voids. At low temperature, energetically accessible local particle configurations can be organized into a random tree with nodes and edges denoting configurations and micro-string propagations respectively. Such trees defined in the configuration space naturally describe systems defined in two- or three-dimensional physical space. A micro-string propagation initiated by a void can facilitate similar motions by other voids via perturbing the random energy landscape, realizing  path interactions between voids or equivalently string interactions. We obtain explicit expressions of the particle diffusion coefficient and a particle return probability. Under our approximation, as temperature decreases,  random trees of energetically accessible configurations exhibit a sequence of percolation transitions in the configuration space, with local regions containing fewer coupled voids entering the non-percolating immobile phase first. Dynamics is dominated by coupled voids of an optimal group size, which increases as temperature decreases. Comparison with a distinguishable-particle lattice model (DPLM) of glass shows very good quantitative agreements using only two adjustable parameters  related to typical energy fluctuations and the interaction range of the micro-strings. 
\footnote{Published as C.-H. Lam, J. Stat. Mech. {\bf 2018}, 023301 (2018).}
\end{abstract}
\maketitle

\section{Introduction}
The nature of the glassy state is one of the most fundamental and long-standing problems in the study of condensed matters \cite{binderbook,ediger2012review,biroli2013review,stillinger2013review}. The mechanism of a dramatic slowdown of the dynamics as the temperature $T$ decreases is highly controversial, despite intensive efforts based on approaches such as the Adam-Gibbs theory \cite{adam1965},  mode-coupling theory \cite{gotzebook}, random first order transition theory \cite{kirkpatrick1989}, dynamic facilitation theory \cite{fredrickson1984,palmer1984,ritort2003review,garrahan2011review} and so on. 

A remarkable feature of glassy dynamics is a string-like particle hopping motion, in which neighboring particles arranged in a line displaces one another. 
It has been found to dominate particle dynamics in many glassy systems as observed in  MD simulations \cite{glotzer1998,glotzer2003,glotzer2004} and experiments 
\cite{weeks2000}. If particles in a string hop simultaneously, the motion is referred to as coherent. A coherent string or a coherent segment in a string is called a micro-string \cite{glotzer2004}. 
Recently, we have conducted MD simulations \cite{lam2015} of a bead-spring model of polymers and observed that string-like motions become highly repetitive at low $T$.
This is directly related to back-and-forth particle hopping motions widely studied \cite{miyagawa1988,vollmayr2004,vogel2008,kawasaki2013,ahn2013,helfferich2014,yu2017}.
A return probability of hopped particles for simulated polymers studied in Ref. \cite{lam2015} reaches 73\% at low $T$ at equilibrium and it is even higher for non-equilibrium samples quenched to even lower $T$. Equilibration of quenched samples show further increase in the return probability. The simulation results indicate that as $T$ decreases the return probability approaches towards unity, implying significantly slowed down dynamics.
Examinations of particle trajectories show that those strings spatially isolated from other strings tend to repeat for much longer durations. 
A string typically breaks free of repetitions via a pair interaction with another string, realizing facilitated dynamics of strings. 

Lattice models also play important roles in the study of glass and are often more tractable analytically \cite{fredrickson1984,palmer1984,kob1993,newman1999,darst2010,lipson2013}.
We have proposed a distinguishable-particle lattice model (DPLM) of glass
\cite{lam2017dplm} to further study the string repetition and string interaction phenomena observed in MD simulations in Ref. \cite{lam2015}. 
The DPLM is a lattice gas model with infinite particle types generalizing an identical-particle sliding block model of glass \cite{palmer1990}. It is closely related to lattice models of glass with particles of a few \cite{darst2010} or many types \cite{sasa2012,rabin2016lattice}, and is also related to many-species molecular models \cite{rabin2015,rabin2016,tanaka2016}. 
Typical glassy dynamics are demonstrated for a wide range of temperature and particle density. 
Defined by a simple energy function without explicit kinetic constraint except for simple particle exclusion, it exhibits emergent facilitation behaviors and is thus suggested to provide microscopic justifications for kinetic constraints assumed in kinetically constrained models (KCM) \cite{fredrickson1984,palmer1984,ritort2003review,garrahan2011review}. In addition, 
DPLM simulations reproduce the convergence of particle return probability towards unity as $T$ decreases, analogous to that observed in MD simulations \cite{lam2015}.
An additional property important to the present work is that equilibrium statistics are exactly known, rendering it as analytically tractable as more coarse-grained and energetically trivial models such as the KCM's. 

In this paper, we describe the dynamics observed in polymer and DPLM simulations based on micro-strings in the presence of a random energy landscape in the particle configuration space. A random-tree theory is proposed  and it is applied to derive particle diffusion coefficient characteristic of liquid and glass at high and low $T$ respectively. We expect that the theory is generally applicable to glassy systems exhibiting string-like motions including polymer and DPLM simulations. Nevertheless, since required model parameters are reliably available only for the DPLM, we compare the theory quantitatively only to DPLM simulations. We demonstrate that using only two  adjustable parameters, analytic results from the random tree theory are in good agreement with DPLM simulation data from Ref. \cite{lam2017dplm}. 

The rest of this paper is organized as follows. \Sec{dplm} explains and further analyzes the DPLM. \Sec{microstring} describes both polymer and DPLM simulations on the same footing based on micro-strings. We formulate a random tree theory for a local region with a single void in \sec{singlevoid}. Our main results are then obtained by generalizations to include void coupling effects in local regions with multiple voids in \sec{mvoid}. Limiting cases for the liquid and glass phases are studied in \sec{limits}. \Sec{conclusion} concludes the paper.

\section{ Distinguishable-particle lattice model (DPLM)}
\label{dplm}
Following Ref. \cite{lam2017dplm}, the DPLM is defined in $d=2$ dimensions with $N$ particles on a $L\times L$ square lattice following periodic boundary conditions. It is an interacting lattice gas model with distinguishable particles. No more than one particle can occupy a site. If site $i$ is occupied, the particle index $s_i(t) = 1,2,\dots, N$ specifies which particle occupies the site at time $t$. Otherwise, $s_i(t)=0$ and the site is occupied by a void.
The system energy at time $t$ reads
\begin{equation}
\label{E}
E = \sum_{<i,j>'} \Vijt
\end{equation} 
where the sum is over occupied nearest neighboring (NN) sites. 

An unique feature of the DPLM is that the disorder is quenched only in the configuration space, but it is $transient$ in the physical space, as is essential for modeling glassy fluids. Specifically, the DPLM involves a site-particle dependent interaction $\Vkl$, which depends on both the sites $i$ and $j$ and the particle indices $k$ and $l$. Each $\Vkl$  follows an $a$ $priori$ distribution $g(V)$ taken as the uniform distribution in $[-0.5, 0.5]$, i.e.
\begin{equation}
  \label{g}
  g(V)=
\begin{dcases}
1           , & \mbox{for~~}  -0.5 \le V \le 0.5  \\ 
0, & \mbox{otherwise}.  \\ 
\end{dcases}
\end{equation}
Since each $\Vkl$ is a quenched random value for any given particles $k$ and $l$ at sites $i$ and $j$, the interaction is fixed for any given local particle configuration. The disorder is therefore quenched in the configuration space. 

Importantly, the quenched variables $\Vkl$ are not involved directly in $E$ in \eq{E}. Instead, they contribute non-trivially via $V_{ij}(t) \equiv \Vijt$.
For any NN sites $i$ and $j$, $\Vijt$ admits an implicit time dependence via $s_i(t)$ and $s_j(t)$, and changes in value when a particle at  $i$ or $j$ is replaced. Hence, the disorder is $not$ quenched in the physical space. In a simulation with $N$ particles, each $\Vijt$ continuously samples over order $N^2$ possible $\Vkl$'s. For large $N$ in the ergodic phase, any occurred value $\Vkl$ in general repeats with a vanishing probability after long time because particles $k$ and $l$ will have diffused far away from sites $i$ and $j$.
Any value $\Vijt=\Vkl$ is thus only of a transient nature. The absence of   disorder quenched in the physical space has been verified in simulations by the vanishing of  the self-intermediate scattering function at long time \cite{lam2017dplm}. The quenched variable $\Vkl$ only leads to quenched disorder in the configuration space. Its role is analogous to quenched variables such as the particle diameter $\sigma_k$ of particle $k$ in a poly-dispersive colloidal liquid.

The dynamics follow standard kinetic Monte Carlo rules. Each particle can hop to an unoccupied NN site at a rate 
\begin{equation}
  \label{w}
    w = w_0 \exp\left(-\frac{E_{0} + \Delta E/2}{k_BT}\right)
\end{equation}
where $\Delta E$ is the change in the system energy $E$ due to the hop. We take $E_{0} = 1.5$ and $w_0 = 10^{6}$ following Ref. \cite{lam2017dplm}. 
Kinetic Monte Carlo simulations of the DPLM using \eq{w} for a wide range of $T$ and void density $\phiv$ show no sign of undesired crystalization or particle segregation. At high $T$ or large $\phiv$, behaviors typical of simple non-glassy fluids are observed. By contrast, at low $T$ and small $\phiv$, one observes typical behaviors of glass including a plateau in the  mean square displacement (MSD), a super-Arrhenius $T$ dependence of the particle diffusion coefficient $D$, a self-intermediate scattering function decaying stretched-exponentially towards zero at long time, a violation of the Stokes-Einstein relation, and an increasing four-point susceptibility as $T$ decreases. 

For small void density $\phi_v$, the particle diffusion coefficient $D$ at any given $T$ exhibits the scaling relation
\begin{equation}
  \label{scaling}
  D \sim \phi_v^\alpha .
\end{equation}
At high $T$, $\alpha \simeq 1$ indicating particle motions induced by independent void motions. As $T$ decreases, 
$\alpha$ rises monotonically. For $\alpha\simeq 2$ for instance, $D \sim \phiv^2$ corresponds to motions dominated by coupled pairs of voids while single voids are trapped. This demonstrates emergent facilitated dynamics of voids.

A feature crucial to this work is that equilibrium statistics of the DPLM is known exactly. 
At any instant, we refer to an interaction $\Vij$ which directly appears in the energy function in \eq{E} as realized, and $\Vkl$ as unrealized if  
particle $k$ and $l$ are not currently located at site $i$ and $j$ respectively. 
At equilibrium, an unrealized interaction $\Vkl$ simply follows the $a$ $priori$ distribution $g(\Vkl)$ in \eq{g}. In contrast, 
a realized interaction $\Vij$ follows a posteriori distribution 
\begin{equation}
  \label{peq}
  p_{eq}(\Vij) = \frac{1}{\mathcal N} e^{-\upbeta \Vij}  g(\Vij)
\end{equation}
where $\mathcal N = \int e^{-\upbeta V} g(V) dV$ is a normalization constant and $\upbeta=1/k_BT$ with $k_B=1$ being the Boltzmann constant. The distribution $g$ is thus analogous to a density of states and it is being  weighted by the Boltzmann factor in \eq{peq}.

In this work, we now further analyze the dynamics of the DPLM theoretically.
We describe a particle hop equivalently as the hop of a void in the opposite direction. For a hop attempt of an isolated void to an occupied NN site, \eq{E} implies an energy change of the system given by
\begin{equation}
  \label{DeltaE}
 \Delta E = \sum_{\gamma=1}^3 (\Vgamma' - \Vgamma) ,
\end{equation}
 where $\Vgamma$ denotes the three realized interactions following $p_{eq}$ to be broken, while $\Vgamma'$ denotes the three prospective interactions following $g$ to be realized. We obtain the probability distribution $P(\Delta E)$ of $\Delta E$ by numerically performing the convolution of the six distributions from \eqs{g}{peq} of these six random variables.

To allow further analysis, we consider a hop attempt energetically possible if 
\begin{equation}
  \label{kT}
 \Delta E \le  \DEmax
\end{equation}
where $\CDE$ is a tunable parameter related to typical energy fluctuations induced by a micro-string and is expected to be of order $1$.
Otherwise, it is deemed  unlikely and neglected. The probability $q$ that an allowed hop is energetically possible follows
\begin{eqnarray}
\label{q}
  q = 
\int_{-\infty}^{\DEmax}P(\Delta E) d \Delta E.
\end{eqnarray}
As $T$ decreases, $q$ decreases monotonically from 1 to 0.

Denote the particle coordination number of the square lattice used in the DPLM by $z_0=4$.  Let $z$ be the number of possible hops an isolated void is allowed to make. Since only nearest neighboring hops are allowed, $z=z_0=4$. However, on average only $zq$ of the $z$ allowed hops are energetically possible. 
At equilibrium, excitation and de-excitation events must be balanced so that $\Delta E$ averages to 0. Most hops following \eq{kT} thus also follows ${\mid\Delta E\mid} \le \DEmax$.
For the energetically possible hops, we then neglect the small energy fluctuations $\Delta E$ in \eq{w} and  the rates are approximated by the uniform rate
\begin{eqnarray}
\label{wbar}
\w = w_0 \exp( - E_0 /k_BT ) .
\end{eqnarray}
This also implies the assumption of a uniform energy for all energetically accessible configurations. Similar to MD and DPLM simulations, the simplified dynamics defined here follows detailed balance.

\section{Micro-strings, voids and distinguishable particles}
\label{microstring}
During a string-like motion, each particle arranged in a line of particles hops to displace the preceding one \cite{glotzer1998}. If particles in a string hop simultaneously, it is referred to as a coherent string or a micro-string \cite{glotzer2004}. A string, typically a longer one, can be incoherent and consists of a number of micro-strings.
The average length of strings typically varies mildly from 1.3 to 2.8 particles as $T$ decreases \cite{douglas2013}. 
Micro-strings are even shorter as they are constituents of strings. The dependence of their average length $\l$ on $T$ hence must also be mild, and  is neglected in this work for simplicity.

In many systems, string-like motions are found to dominate relaxation dynamics \cite{glotzer1998,glotzer2003,glotzer2004,weeks2000}. Micro-strings, being their constituting components, thus also dominate relaxation dynamics.
The coherent hops of particles in a micro-string can be energetically preferred \cite{swayamjyoti2014} because particle bonds within the micro-string need not be broken. 
The structural feature allowing a micro-string motion has been proposed as an elementary excitation in glassy systems \cite{chandler2011}. 
On the other hand, free volumes are long known to be important for glassy dynamics \cite{cohen1961}. 
Motivated by the observation of highly repetitive particle hops dominated by hopping distances comparable to the particle diameter, we have suggested that the existence of a void of a free volume comparable to that of a particle is the elementary excitation allowing a micro-string  \cite{lam2015}. 
Motions of free volumes typically 95\% of the small particle size have also been inferred from experiments on non-equilibrium  glassy colloidal liquids following a bimodal particle radius distribution \cite{han2017}.
In this paper, we follow Ref. \cite{lam2015} and assume that a micro-string must involve a void and the propagation of a micro-string  transport its void from one end of the micro-string to the other.  
More precisely, such a void is a quasi-particle transported by a micro-string in whole, but can be fragmented and distributed among a few interstitial volumes before and after a hop to lower the free energy. This may reconcile with the fact that contiguous voids or free volumes are small and few, and have been found to show only small correlations with particle motions \cite{starr2002,weitz2005,harrowell2006}.  

For the DPLM, the above picture of particle motions based on voids and micro-strings is also applicable. Voids exist and participate in particle motions by definition. Since particles perform independent single hops, each  hop constitutes a micro-string of unit length implying an average micro-string length of $\l=1$ particle.  

Conversely, quantitative measures developed for the DPLM in \sec{dplm} can also be generalized to molecular systems with dynamics dominated by string-like motions. Specifically,   
we define $q$ as the probability that a geometrically allowed micro-string is energetically possible so that it varies from 1 to 0 as $T$ decreases.
Also, we define $zq$ as the average number of energetically possible micro-strings executable by a void. 
Then, $z$ can be interpreted as the average number of geometrically allowed micro-strings, which we limit to within a commonly observed micro-string length, say up to about 3 particles long \cite{glotzer2004}. It may then relate to  the particle coordination number $z_0$ in the physical space very roughly as $z \sim z_0^3$.  
Analogous to \eq{wbar}, we also assume in general for simplicity that all energetically possible micro-strings have a uniform propagation rate $\w$. This effectively also assumes a uniform energy of all configurations in the random tree. 

Throughout this work, we will consider distinguishable particles. For the DPLM, particles  are distinguishable by definition. For molecular systems, examples like polydisperse hard disks and monomers in polymer chains also admit distinct particle properties in general and this justifies the  distinguishability. For other glasses especially monoatomic ones, it is less trivial. It may describe the fact that different particles, upon hopping into a void, in general lead to different frustrations and hence different interparticle distances and interaction energies, analogous to the distinguishable particle case.

\section{Random tree theory for Isolated voids}
\label{singlevoid}

\begin{figure}[tb]
 \includegraphics[width=\columnwidth]{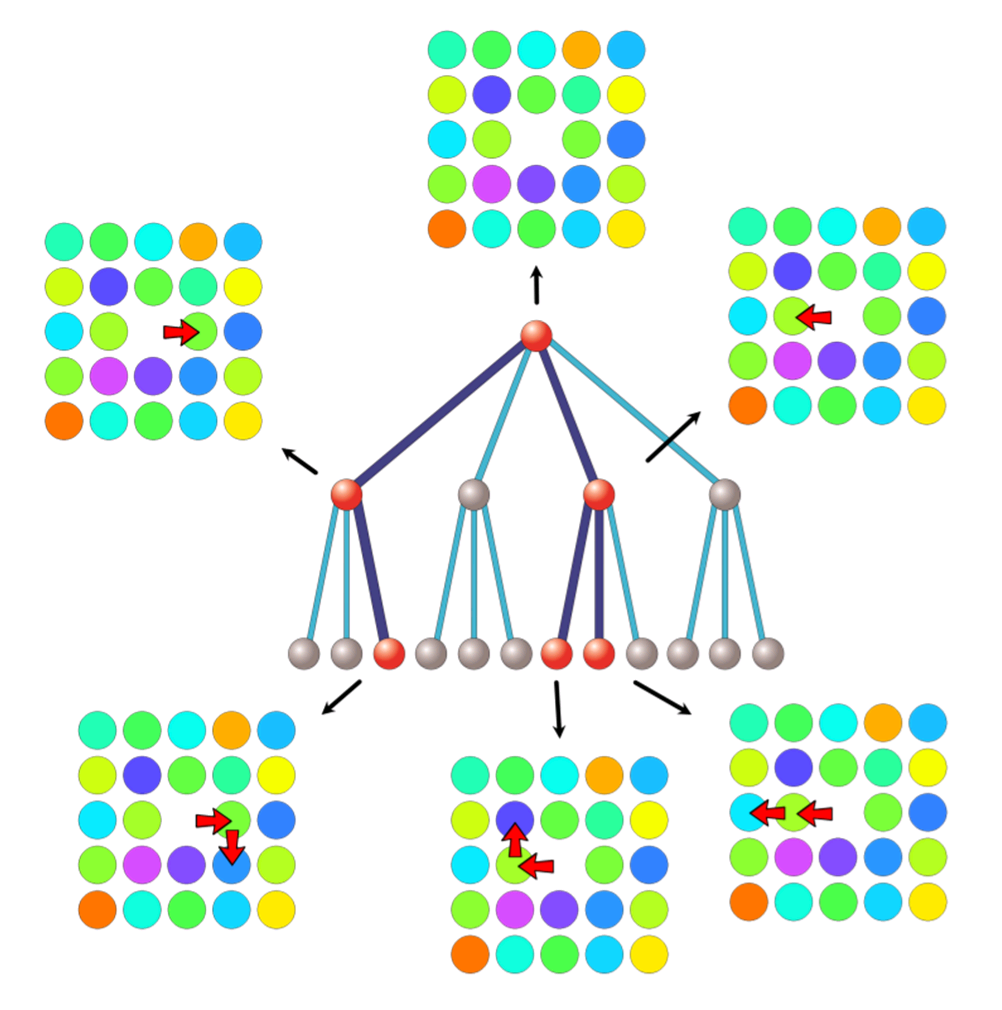}
\caption{The first three levels of a graph in the form of a tree or a Bethe lattice representing the configuration space of a local region with a single void in the DPLM. Each node denotes a particle configuration of the region. An edge joins two configurations connected by an allowed micro-string propagation, corresponding to a hop of a particle or equivalently of a void. The node coordination number $z$ of the Bethe lattice equals the particle coordination number $z_0=4$ of the square lattice representing the physical space. Starting from a local particle configuration associated with the root, void motions (red arrows) leading to 5 examples of descendent nodes are illustrated. Particles are randomly colored to highlight their distinguishability. 
The propagation of a micro-string may be energetically possible (dark blue) with a probability $q$. The energetically accessible nodes (red spheres) form a local random  configuration tree.
The random tree shown is illustrated for e.g. $q=0.5$ so that its average degree is  $c_1=1.5$. 
}
\label{Figtree}
\end{figure}

In this and the next session, we develop a theory of glassy dynamics based on micro-string motions applicable to both molecular systems and the DPLM.
Consider a $d$-dimensional physical space, where $d$=2 or 3. Volumes in the followings are implied to be $d$-dimensional volumes. Let $\phi$ and $\phiv$ be the densities of particles and voids respectively. The average volume $\Omega$ of a particle and of a void follows 
\begin{eqnarray}
\label{Omega}
1/\Omega= \phi+\phiv.
\end{eqnarray} 

Consider a local region of volume $\Vol$ with a single isolated void. 
More details about the definition of such a region will be discussed in \sec{mvoid}.
A local configuration specifies the positions of all distinguishable particles in the region.
The set of all possible configurations constitutes the configuration space, which
can be organized as a graph \cite{newmanbook} by defining nodes as  configurations and edges connecting pairs of configurations related by allowed micro-string propagations. 
\Fig{Figtree} shows an example of such a configuration graph for the DPLM describing a $5\times 5$ region with a single void. The graph has been further arranged as a tree and only the first three levels are shown. The root, defined as level 0, is statistically equivalent to any other nodes and can be used to represent any given configuration, after neglecting boundary effects of the region. In this example, it is associated with a configuration with the void at the center. It is directly connected to 4 nodes in the first level and further to 12 nodes at the second level corresponding to all possible single and double hops of the void respectively. 

\begin{figure}[tb]
 \includegraphics[width={0.75\columnwidth}]{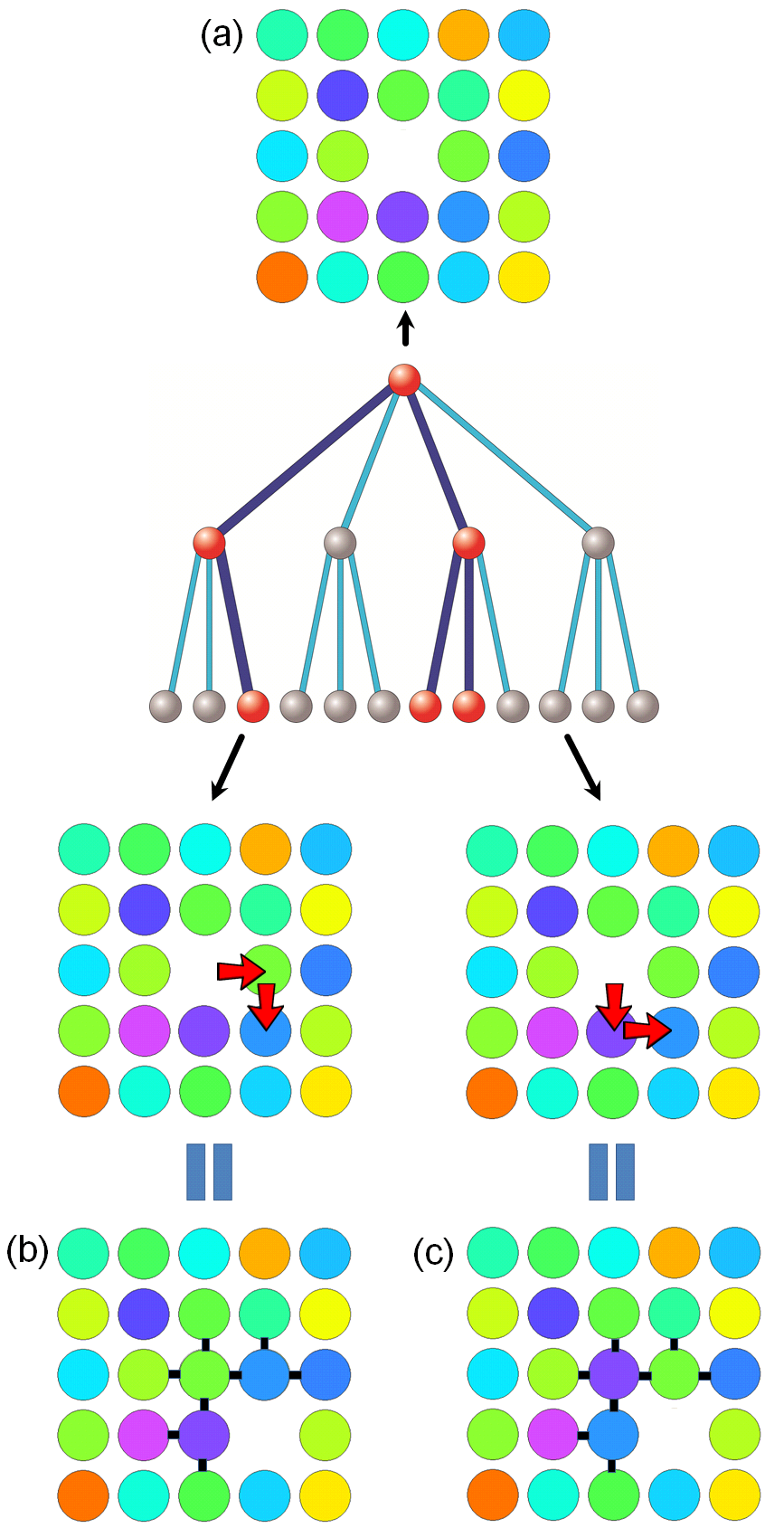}
\caption{The same Bethe lattice and random tree for the DPLM as in \fig{Figtree}. Starting from configuration (a) corresponding to the root, two hops of the void in the clockwise direction lead to configuration (b). Alternatively, two hops in the anti-clockwise direction lead to a different configuration (c), with 3 particles located differently and 8 interactions different (black lines) despite an identical void position. 
}
\label{Figtreeloop}
\end{figure}

As explained in \sec{microstring}, we assume particle distinguishability. A configuration thus specifies not only the position of the void but also the positions of all particles. An important consequence is illustrated in \fig{Figtreeloop} for the same DPLM example, in which configurations (b) and (c) have the same void position. For identical particle systems, they represent identical configurations with an equal energy. The two corresponding nodes are then identical and can be merged together leading to a loop. The tree geometry is then a very rough  approximation.
In sharp contrast, due to particle distinguishability, configurations (b) and (c) are different because 3 particles are located differently. In fact, all 12 nodes in the second level are distinct and the tree structure represents the exact geometry of the graph so far.

Each node in the configuration graph is directly connected to $z$ edges, where $z$ denotes the number of allowed micro-strings executable by a void as explained in \sec{microstring}, Neglecting loops, the configuration graph takes the geometry of a Bethe lattice with a node coordination number $z$. For the DPLM, the Bethe lattice exactly represents the configuration graph for up to the 5th tree-level, only beyond which loops occur. 
This is because for a $2\times 2$ block with a void and 3 particles, 6 consecutive hops of the void in the clockwise direction is the least required to arrive at the same configuration resulting from 6 analogous hops of the void in the anti-clockwise direction. 
Equivalently, 12 consecutive hops of the void return the region to the original configuration as each of the 3 particles would have hopped 4 times and returned to their original positions. The Bethe lattice is thus a very good approximation for the single void case. We emphasize that the Bethe lattice here approximates the high-dimensional configuration space. This can be  drastically more reliable than approximating the low-dimensional physical space by the Bethe lattice, as is much more commonly considered.

It is tempting to consider the dynamics as a random walk of the void in the $d$-dimensional physical space with a random energy landscape. However, this is not appropriate for distinguishable-particle systems, and most likely also for glass in general, because the void position does not uniquely determine the system energy $E$. Using the example of the DPLM in \fig{Figtreeloop} again, configurations (b) and (c) have the same void position but are in fact distinct. Their energies are in general different with eight of the interactions $\Vij$ (indicated by short black lines) in configuration (b) distinct from those in configuration (c). For distinguishable-particle systems in general, the system energy $E$ depends on the detailed particle configurations. The dynamics is thus a random walk of the configuration in a random energy landscape defined in the configuration space rather than in the $d$-dimensional physical space.

Denote the current particle configuration by the root of the Bethe lattice without loss of generality.  
We classify allowed micro-strings as energetically possible or impossible with probabilities $q$ and $1-q$ respectively 
as explained in \sec{microstring}.
The root is then only connected to a random sub-tree of the Bethe lattice separated from other nodes by idealized insurmountable energy barriers. An example of a random tree is illustrated in \fig{Figtree}. 
This local random configuration tree of energetically accessible nodes is our main focus of study. The average number of children of each node is 
\begin{eqnarray}
\label{c1}
c_1 = (z-1)q ,
\end{eqnarray}
except for the root which has on average $zq$ children. Here, $c_1$ is called the average degree, noting that the tree is approximately an Erd\"os-R\'enyi random graph, also called a Poisson random graph \cite{newmanbook}.

As $T$ decreases, $q$ decreases monotonically from 1 to 0 as explained in \sec{microstring} and \eq{c1} implies that $c_1$ decreases from $z-1$ to 0. Noting that $z=4$ for the DPLM and it is expected to be larger for molecular systems, we have $z\gg 1$ in general. There must exist a temperature $T_1$ at which  $q=1/(z-1)$ so that $c_1=1$. For $T<T_1$ so that $c_1<1$, it is straightforward to show that the random tree must be finite with on average $N_{tree}=(1-q)/(1-c_1)$ nodes. 
For $T\ge T_1$ so that $c_1 \ge1$, the tree can be infinite and the  average number of nodes $N_{tree}$ diverges, i.e.
\begin{equation}
  \label{Ntree}
  N_{tree}=
\begin{dcases}
\infty           , & \mbox{for~~} c_1\ge 1  \\ 
\frac{1-q}{1-c_1}, & \mbox{for~~} 0<c_1<1.  \\ 
\end{dcases}
\end{equation}

To describe the dynamics of a local region with a single void, we will consider the following equivalent pictures:
\begin{enumerate}
\item random walk of the configuration in the random tree, characterized by the net level transcending rate $\mu$;
\item random walk of the void in the physical space, characterized by the net hopping rate $\Rv$ of the void;
\item random walks of the particles in the physical space, characterized by the particle diffusion coefficient $D_1$.
\end{enumerate}
Below, We will calculate $\mu$, $\Rv$ and hence $D_1$. 
 
First, we study the random walk of the configuration in the random tree.
Note that all edges corresponding to energetically possible micro-string propagations are assumed the uniform rate $\w$ for simplicity as explained in \sec{microstring}. The quenched disorder in the configuration space is thus encoded only as kinetic constraints on the Bethe lattice and realized as the geometry of an energetically trivial random tree. These simplifications lead to a simplified system analogous to the KCM's \cite{garrahan2011review}.

The random walk of the configuration is thus a simple unbiased random walk in the random tree.
For $T \ge T_1$ so that $c_1 \ge 1$, the random tree can be infinite according to \eq{Ntree}.  
Beyond the root, the rate of a level-decreasing hop is $\w$, while that of a level-increasing hop is on average $c_1\w$. Neglecting the different rate at the root which is irrelevant at long time, the level transcending rate, defined as the net rate of increase in the level, is
$\mu=(c_1-1)\w$ \cite{cassi1989}. 
For $T < T_1$ so that $c_1 < 1$, the tree is finite according to \eq{Ntree}.
On average, the level first increases from 0 before being saturated with $\mu=0$.
At long time, there are indefinite recurrence of the $N_{tree}$ configurations and indefinite back-and-forth repetitions of the $N_{tree}-1$ micro-strings.  Summarizing, we have 
\begin{equation}
\label{mu}
\mu=
\begin{dcases}
(c_1-1)\w           , & \mbox{for~~} c_1\ge 1    .\\
0,                    & \mbox{for~~} 0<c_1<1 \\ 
\end{dcases}
\end{equation}
This shows an mobile-to-immobile transition at $c_1=1$, corresponding to a percolation transition of the random tree in the configuration space. The non-analytic behavior at $c_1=1$ will be inherent by further results to be derived. It is only an artifact of the long-time and other approximations and will be further discussed in \App{nonanalytic}. 

    Second, we project the random walk of the configuration into a random walk of the void in the physical space. The tree-level transcending rate $\mu$ simply equals the net micro-string propagation rate. 
Noting that each micro-string on average involves $\l$ correlated hops by a void,  
the net hopping rate of the void is thus $\Rv =\l\mu$.
Applying \eq{mu}, we have
\begin{equation}
\label{Rvoid}
\Rv=
\begin{dcases}
\l(c_1-1)\w        , & \mbox{for~~} c_1\ge 1    .\\
0,                   & \mbox{for~~} 0<c_1<1 \\ 
\end{dcases}
\end{equation}
We emphasize that $\Rv$ calculated here is a net hopping rate accounting for the non-reversal part of the random walk of the void. The canceling effects of the back-and-forth motions are already accounted for and their slow-down effects are reflected by the factor $c_1-1$ for $c_1\ge 1$ and by $\Rv=0$ otherwise.
The net motion of the void is a non-reversal random walk in the physical space which forbids backward retracing of its trajectory.
Nevertheless, it is not a self-avoiding walk. For example, the void can perform a loop and return to a previously visited position.  

    Finally, we study the motions of the particles as induced by those of the  void. There are on average $\phi \Vol$ particles in the local region. Since each hop of the void is equivalent to a hop of a particle in the opposite direction, the net hopping rate per particle in the region is 
\begin{eqnarray}
\label{Ralpha}
R_{ptcle} = \Rv  / \phi \Vol .
\end{eqnarray}
Similarly, $R_{ptcle}$ is also a net rate characterizing the non-reversal part of the particle random walk. A subtle point is that the non-reversal property here forbids reversing the evolution of the full particle configuration in the local region, rather than the trajectory of an individual particle. The back-and-forth motions of the particles due to those of the void are thus accounted for. Successive steps of the net motion of each particle are expected to be only slightly correlated.
Note that not all reversal steps of particles are excluded. For example, the void can first induce a hop of a particle, perform a loop in the physical space, and then return to induce a second hop of the same particle in the opposite direction. This is not a reversal step of the void nor of the local particle configuration and it is not excluded. The net motion of a particle is thus neither a non-reversal nor a self-avoiding random walk. 

We approximate the net motions of the particles by simple uncorrelated random walks. This is expected to be a good approximation because the strongest correlations in the form of back-and-forth motions of the particles induced by those of the configuration have been subtracted out. 
The particle diffusion coefficient in a single-void region is then $D_1 = a^2 R_{ptcle}/(2d)$, where $a$ is the average particle hopping distance and is comparable to the particle diameter. Applying also Eqs. (\ref{Rvoid}), (\ref{Ralpha}) and (\ref{Omega}), we get
\begin{eqnarray}
\label{D1}
D_1 = 
\begin{dcases}
 \frac{(c_1-1) \l \w a^2 \Omega }{2d (1-\phi_v\Omega)\Vol }, & \mbox{for~~} c_1\ge 1    \\
0,                    & \mbox{for~~} 0<c_1<1 \\ 
\end{dcases}
\end{eqnarray}

    A dependence of $D_1$ on $\Vol$ may appear unexpected for this single-void case. This is because the  physically relevant quantity is instead the particle diffusion coefficient $\tilde D$ averaged over all regions. The probability that a region has a single void for small $\phiv\Vol$ can be denoted by $f(1,\phiv\Vol)$, which follows $f(1,\phiv\Vol)\simeq\phiv \Vol$. Regions with no void have null contribution to diffusion. Neglecting in this section regions with multiple voids, the particle diffusion coefficient $\tilde D$ averaged over all regions is hence $\tilde D = f(1,\phiv\Vol) D_1$, which simplifies to
\begin{eqnarray}
\label{Disolated}
\tilde D = 
\begin{dcases}
 \frac{(c_1-1) \l \w a^2 \phiv\Omega }{2d (1-\phi_v\Omega)}, & \mbox{for~~} c_1\ge 1    \\
0,                    & \mbox{for~~} 0<c_1<1 \\ 
\end{dcases}
\end{eqnarray}
and is independent of $\Vol$.

\begin{figure}[tb]
 \includegraphics[width=\columnwidth]{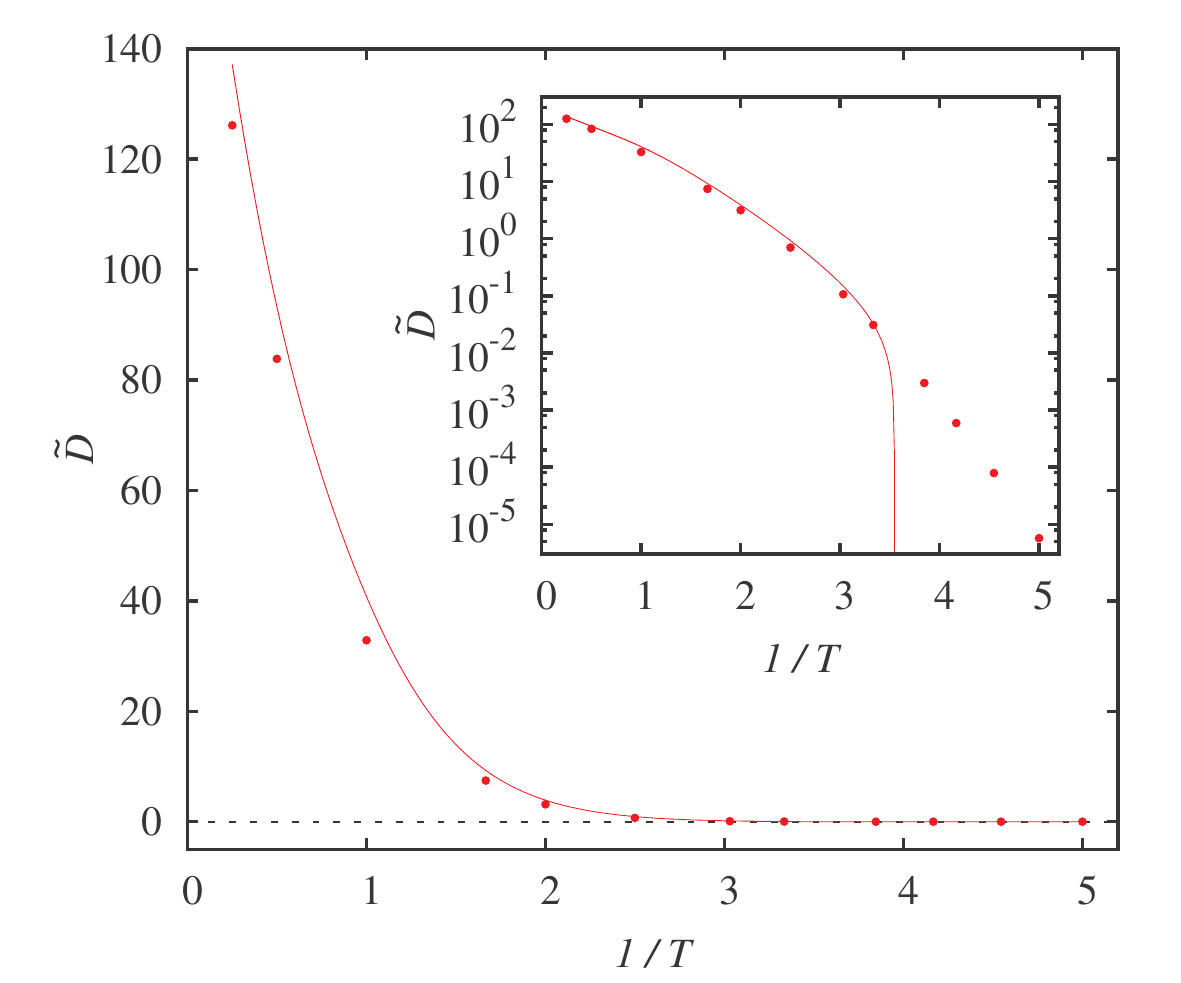}
\caption{
Particle diffusion coefficient $\tilde D$ against $1/T$ for DPLM with a single void from kinetic Monte Carlo simulations (symbols) and random-tree theoretical results in \eq{Disolated}  with a single tunable parameter $\CDE=1.7$ (line). The inset shows the same plot on a semi-log scale.  The discrepancies at low $T$ have vanishing magnitudes and implicate no issue for our main results on the case with multiple voids because of the presence of other additive terms.
}
\label{FigD1}
\end{figure}

\Fig{FigD1} compares \eq{Disolated} with DPLM simulations using small systems of size $ 50 \times 50$ with a single void. Simulation details follow those in Ref. \cite{lam2017dplm} and will be reported elsewhere \cite{anakinunpub}. 
 \Eq{Disolated} evaluated using \eqs{q}{c1} fits the simulation data very well with $\CDE = 1.7$ as the single tunable parameter, after taking $d=2$, $z=4$,  $a=\Omega=\l=1$ and $\phiv=1/50^2$. This choice of $\CDE = 1.7$ indeed best fits the more extensive data to be discussed \sec{mvoid} and its applicability here provides a consistency check of our theory.
 In particular, \fig{FigD1} shows that the simulation data is in approximate agreement with a transition to $\tilde D=0$ as $T$ decreases as predicted in  \eq{Disolated}. 
Nevertheless, the sharp transition in  \eq{Disolated} is only an artifact to be explained in \App{nonanalytic}. In the semi-log plot in the inset of \fig{FigD1}, we observe that the simulation data indeed show a smooth crossover towards $\tilde D =0$. 
The discrepancy at low $T$ signifies that energetically unlikely hops, which violate \eq{kT} and are neglected in our theory, have become important. This will be further discussed in  \App{nonanalytic}. However, it has no impact on our main results to be derived next because in the presence of multiple voids, other additive terms dominate at low $T$.

\section{Random tree theory for Interacting voids}
\label{mvoid}
\begin{figure}[tb]
\begin{flushleft}~~~(a)\end{flushleft} \vspace{-3mm}
\includegraphics[width=0.45\columnwidth]{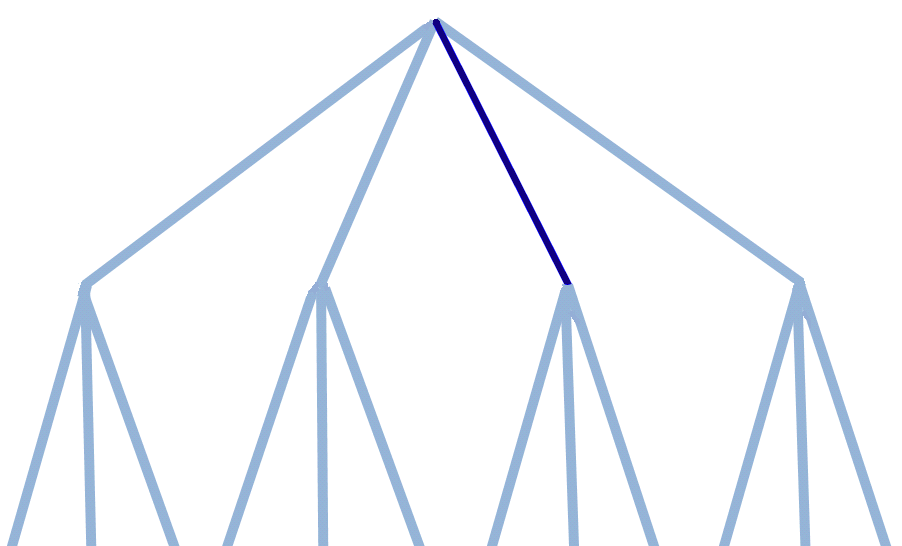}
~
\\~
~
\begin{flushleft}~~~(b)\end{flushleft} \vspace{-3mm}
\includegraphics[width=0.9\columnwidth]{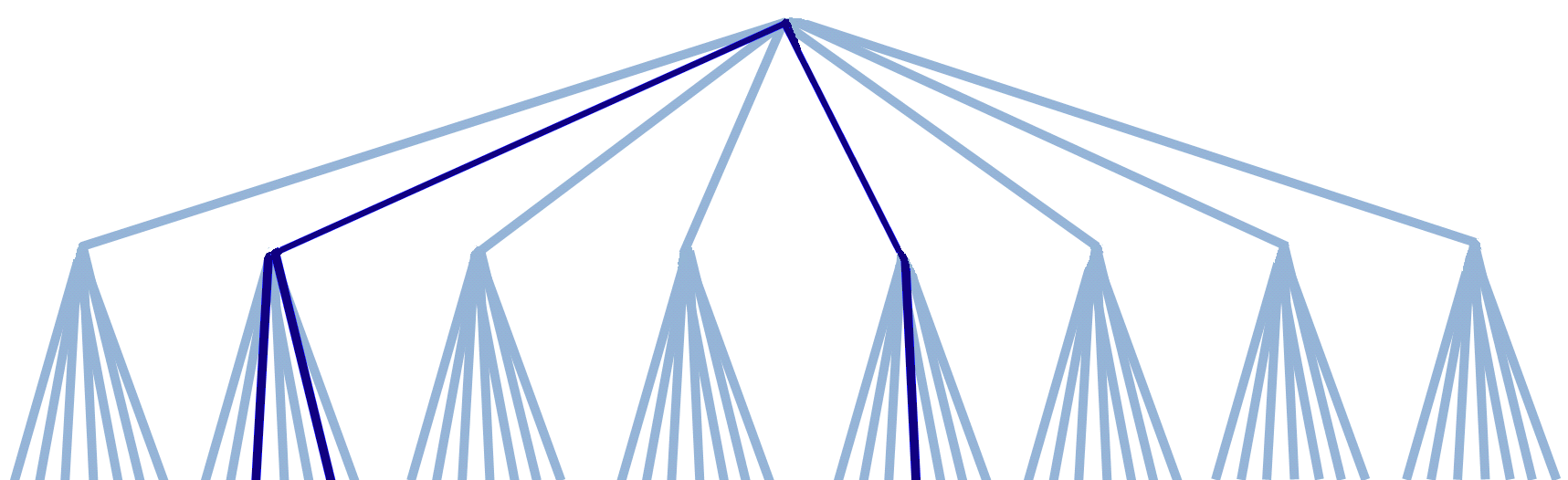}
\caption{(a) A simplified schematic diagram of the first three levels of a Bethe lattice (blue) and a random tree (dark blue) for a local region in the DPLM with a single void. The random tree is illustrated for, e.g. $q=0.25$, so that the average degree is $c_1=0.75$ implying a finite tree. (b) A similar diagram for a local region with two voids illustrated also for $q=0.25$, implying the same $T$. The average degree increases to $c_2=1.75$ and the random tree can be infinite. 
}
\label{Figtree2void}
\end{figure}

We have considered a single void associated with $z$ allowed micro-strings, leading to a node coordination number $z$ of the Bethe lattice representing the configuration space.   
\Fig{Figtree2void}(a) shows a schematic diagram of the Bethe lattice for a region with $m=1$ void and $z=4$, which is a simplified version of that in \fig{Figtree}. 

We now generalize our results to include local regions with $m \ge 2$ voids. There are in general $mz$ allowed ways that the local region can evolve via a single micro-string propagation by any of the $m$ voids. After one of these micro-string propagates, there are again $mz$ allowed micro-string propagations including the reversed one. A key issue is that some of these $mz-1$ non-reversing micro-strings may not be new and this will be further discussed.
Formally, we can still organize the configuration space into a Bethe lattice with a node coordination number $mz$, as illustrated for example in \Fig{Figtree2void}(b) for the DPLM with $m=2$ and $mz=8$. 

More precisely, some nodes in this Bethe lattice are identical, which upon merging lead to loops starting from the second level. For example, two voids far apart can each initiate a micro-string in arbitrary order and arrive at the same particle configuration, leading to two identical nodes at the second level. We adopt the notation of not merging these nodes. The traversabilities of edges may then admit considerable correlations.

\begin{figure}[b]
\includegraphics[width=0.5\columnwidth]{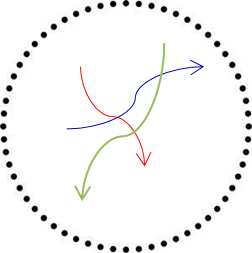}
\caption{A schematic diagram of a local region with 3 voids and 3 associated micro-strings which fully interact. The arrows show the pathways of voids during the propagation of the micro-strings. Due to the spatial overlaps, the propagation of any one of the micro-strings changes the energy landscape experienced by the other voids. The other two micro-strings are in general suppressed and on average two new micro-strings (not shown) are enabled. This enriches the configurations accessible by the dynamics and facilitates motions.
}
\label{Figint}
\end{figure}

To minimize these correlations among edges, a region should have a small volume $\Vol$ so that voids within a region must be closeby to each other.  
Then, we can take the simple approximation that all voids in a region fully interact, i.e. every pair of micro-strings overlap spatially \cite{lam2015}. 
\Fig{Figint} shows schematically a fully interacting set of voids and associated energetically possible micro-strings for $m=3$.  Denoting the current configuration by the root, the 3 micro-strings connects the root to 3 first-level nodes. 
The propagation of any of the micro-strings alters the particle configuration and the energy landscape so that the other two micro-strings are in general suppressed. Under equilibrium conditions, two new energetically possible micro-strings on average are generated, leading to two  distinct second-level nodes for each first-level node. 
This fully interacting scenario reduces the correlations among the edges of the Bethe lattice.
We further assume that each edge in the Bethe lattice is independent of the others and is traversable with a probability $q$. These approximations are further explained using the example of the DPLM in \App{independentrate}.

Applying the above approximations, the energetically accessible configurations then form a random subtree of the Bethe lattice with an average degree $c_m=(mz-1)q$. Applying \eq{c1}, we can write $c_m= m c_1 + ((m-1)/(z-1)) c_1$. Noting $z\gg 1$ for systems of interest, the second term is small. To simplify further algebra, we approximate $c_m$ as 
\begin{eqnarray}
\label{cm}
c_m = m c_1 = m (z-1)q.
\end{eqnarray}
Generalizing \eq{D1} straightforwardly, the particle diffusion coefficient $D_m$ in a region with $m\ge 1$ voids follows
\begin{eqnarray}
\label{Dm}
D_m = \begin{dcases}
  \frac{(c_m-1) \l \w a^2 \Omega }{2d (1-\phi_v\Omega)\Vol }, ~~&  \mbox{for $c_m\ge1$}\\
0,                    & \mbox{for~~} 0<c_m<1 .\\ 
\end{dcases}
\end{eqnarray}
An important consequence is that a larger group size $m$ leads to larger $c_m$ and $D_m$ corresponding to a higher mobility. Denoting $T$ at which $c_m=1$ by $T=T_m$.
For example for $T_2 < T < T_1$, one gets $c_2>1 > c_1$ so that a region with a single-void is in the immobile phase described by a finite random tree while a region with two voids is in the mobile phase described by an infinite random tree, as illustrated by Fig. \ref{Figtree2void}(a) and (b) respectively. 

Finally, voids are expected to have a uniform random spatial distribution,   if aggregation is insignificant such as for the DPLM studied in Ref. \cite{lam2017dplm}. Then, the number of voids $m$ in a region of volume $\Vol$ 
follows a probability distribution $f(m;\barm)$ which can be approximated by the Poisson distribution 
\begin{equation}
\label{Poisson}
f(m;\barm) = \frac{{\barm^m e^{- \barm }}}{m!} .
\end{equation}
with an average $\barm = \phiv \Vol$. 
Averaging over all regions, the particle diffusion coefficient is 
\begin{eqnarray}
\label{D0}
D= \sum_{m=1}^\infty  f(m;\barmv) D_m .
\end{eqnarray}
Using also Eqs. \refb{cm} and \refb{Dm}, we finally get
\begin{eqnarray}
\label{D}
D= \sum_{m>1/c_1}  f(m;\barmv)   \frac{(mc_1-1) \l \w a^2 \Omega }{2d (1-\phi_v\Omega)\Vol } .
\end{eqnarray}
\Eq{D} provides an explicit expression of $D$ and is the main result of this work.
It is non-analytic with respect to $T$ when $c_m=1$ resulting from the percolation transitions of the random trees for various regions. This will further be discussed in \App{nonanalytic}.

By using \eq{Poisson}, we have implicitly assumed the simplest scenario that the physical space is statically partitioned into local regions each of volume $\Vol$. More generally, we envision instead
time-dependent local regions each of which moves and deforms dynamically following the void motions to encapsulate coupled voids and to exclude farther ones not interacting with the group. This is motivated by the observation that a pair of voids can appear bounded while migrating together for an extended period of time in DPLM simulations \cite{lam2017dplm}. This picture of dynamic local regions should induce some quantitative deviation from \eq{Poisson} but should not alter our results qualitatively.
From time to time, voids are exchanged among interacting groups and thus also among the dynamic local regions. We observe from DPLM simulations that this most often occurs when a mobile group of $m$ particles with $c_m>1$ diffuses collectively around,  picking up or dropping down individual voids. 
Both local-region dynamics and void exchanges occur at longer time scales and should have only small quantitative impacts on results in this work. They dictate the residence times of particles in a mobile group and hence the long time dynamics of individual particles, but not the equilibrium population of the mobile groups and hence average quantities such as $D$. Other minor impacts are outlined in \App{nonanalytic}.

\begin{figure}[tb]
 \includegraphics[width=\columnwidth]{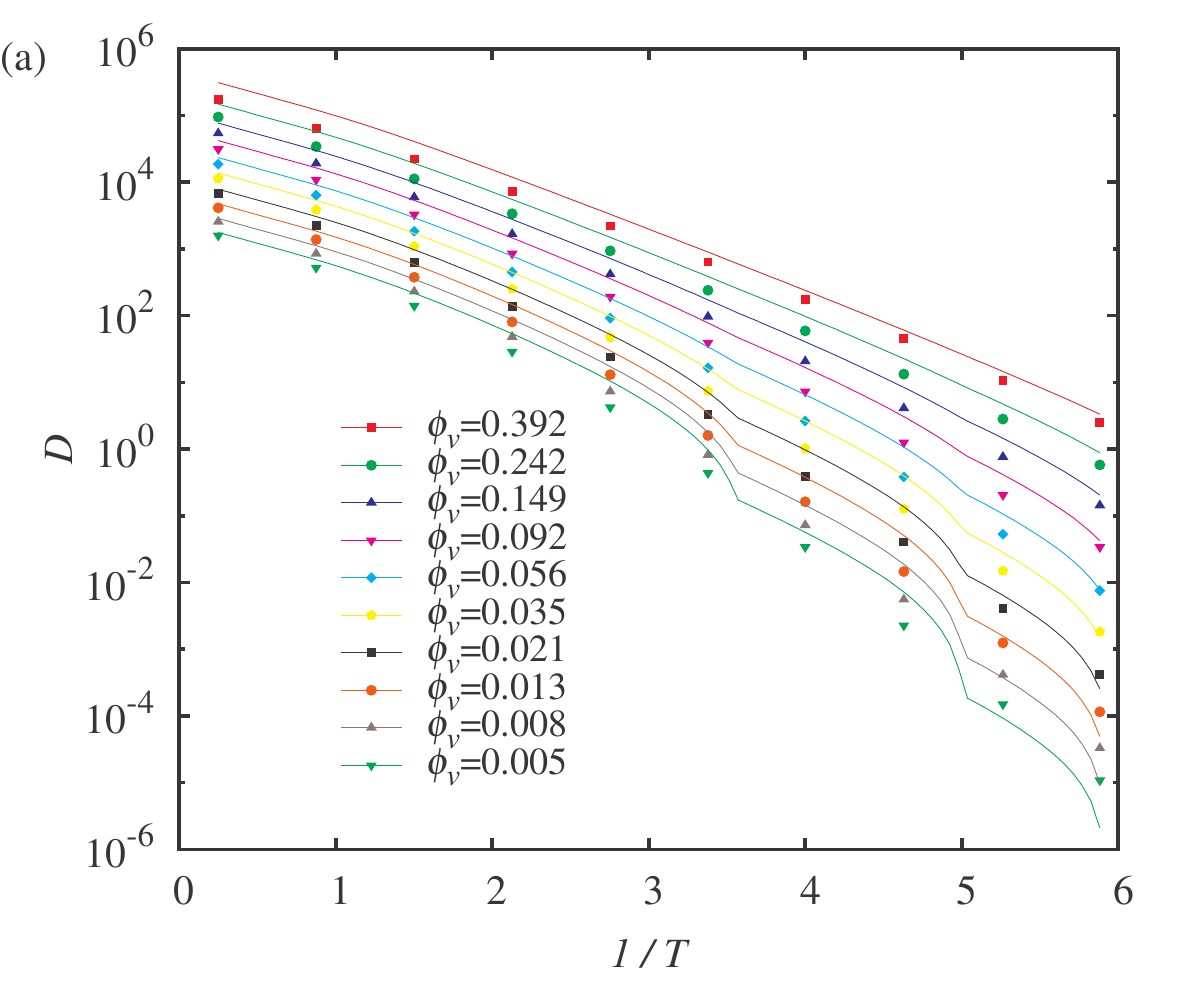}
 \includegraphics[width=\columnwidth]{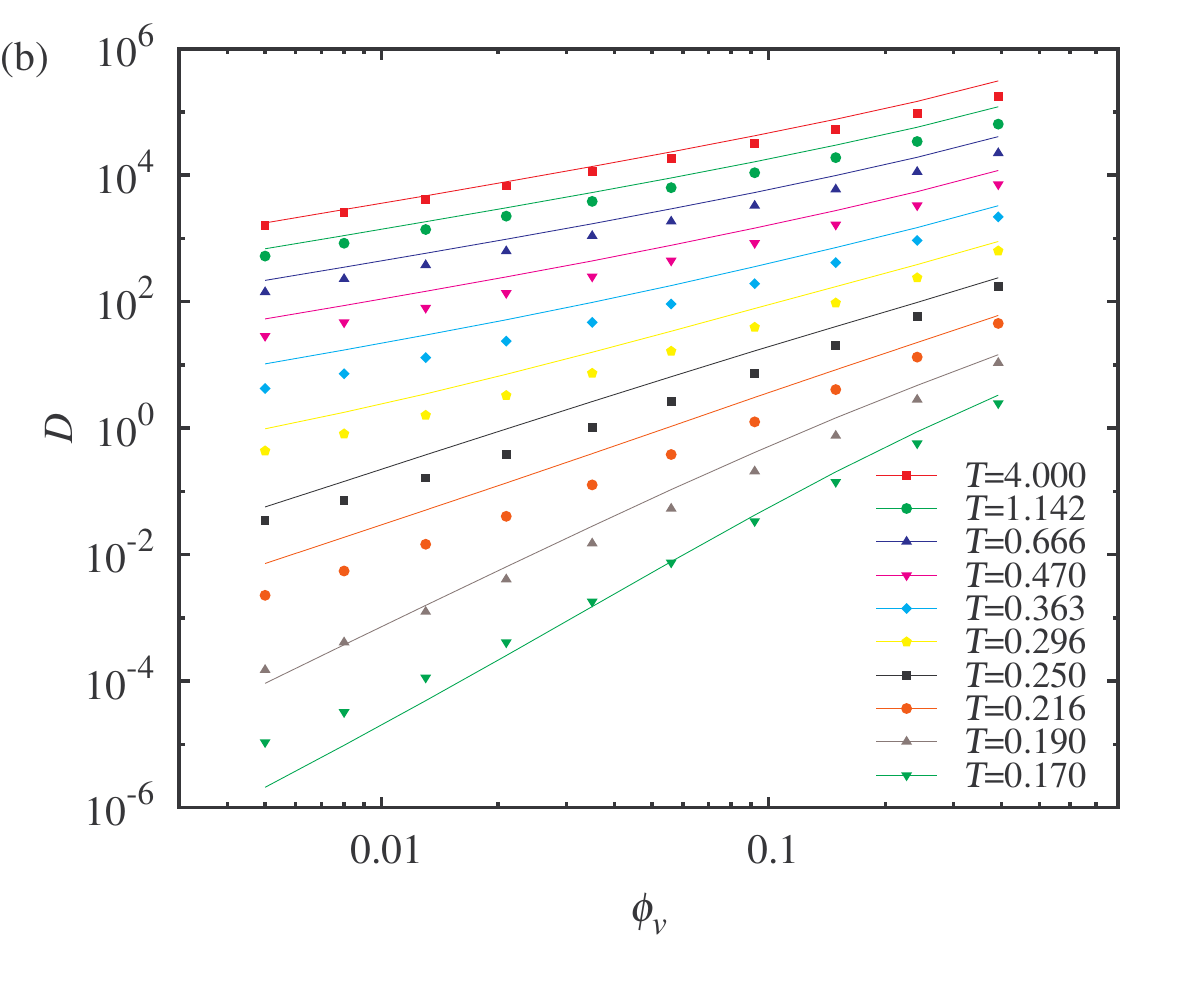}
\caption{Particle diffusion coefficient $D$ against $1/T$ (a) and void density $\phiv$ (b) for the DPLM from kinetic Monte Carlo simulations in Ref. \cite{lam2017dplm} (symbols) and random-tree theoretical result in \eq{D} (lines). The random-tree theory uses $\CDE=1.7$ and $\Vol=12$, which are the only adjustable parameters.}
\label{FigD}
\end{figure}

\Eq{D} will now be checked against kinetic Monte Carlo simulations of the DPLM from Ref. \cite{lam2017dplm}. 
Values of $D$ measured from simulations are plotted as symbols in \Fig{FigD}(a) against $1/T$ for various $\phi_v$. \Fig{FigD}(b) replots the same data set in the form of $D$ against $\phi_v$ for various $T$. To apply our analytic expression in \eq{D} using also \eq{c1}, note that most of the required parameters are exactly known: $d=2$, $z=4$, $a=\Omega=\l=1$, while $q$ is numerically calculated from \eq{q}.
The only fitted parameters are $\CDE=1.7$ and $\Vol=12$. 
Using these parameters, fitted values of $D$ calculated from \eq{D} are plotted as lines in Figs. \ref{FigD}(a) and (b).

The good fits in Figs. \ref{FigD}(a) and (b) show that our analytic theory agrees very well with kinetic Monte Carlo simulations. In particular, two important qualitative features found in the simulations in Ref. \cite{lam2017dplm} are successfully reproduced. First, from Fig. \ref{FigD}(a), $D$ from \eq{D} exhibits a crossover from an Arrhenius $T$ dependence at large $\phiv$ to a super-Arrhenius one at low $\phiv$. The non-analytic behavior at low $\phiv$ will be discussed in \App{nonanalytic}. Second, from Fig. \ref{FigD}(b), the linear regions of the theoretical lines in the log-log plot  for small $\phiv$ reproduce the power-law in \eq{scaling}. 

\begin{figure}[tb]
\includegraphics[width=0.49\columnwidth]{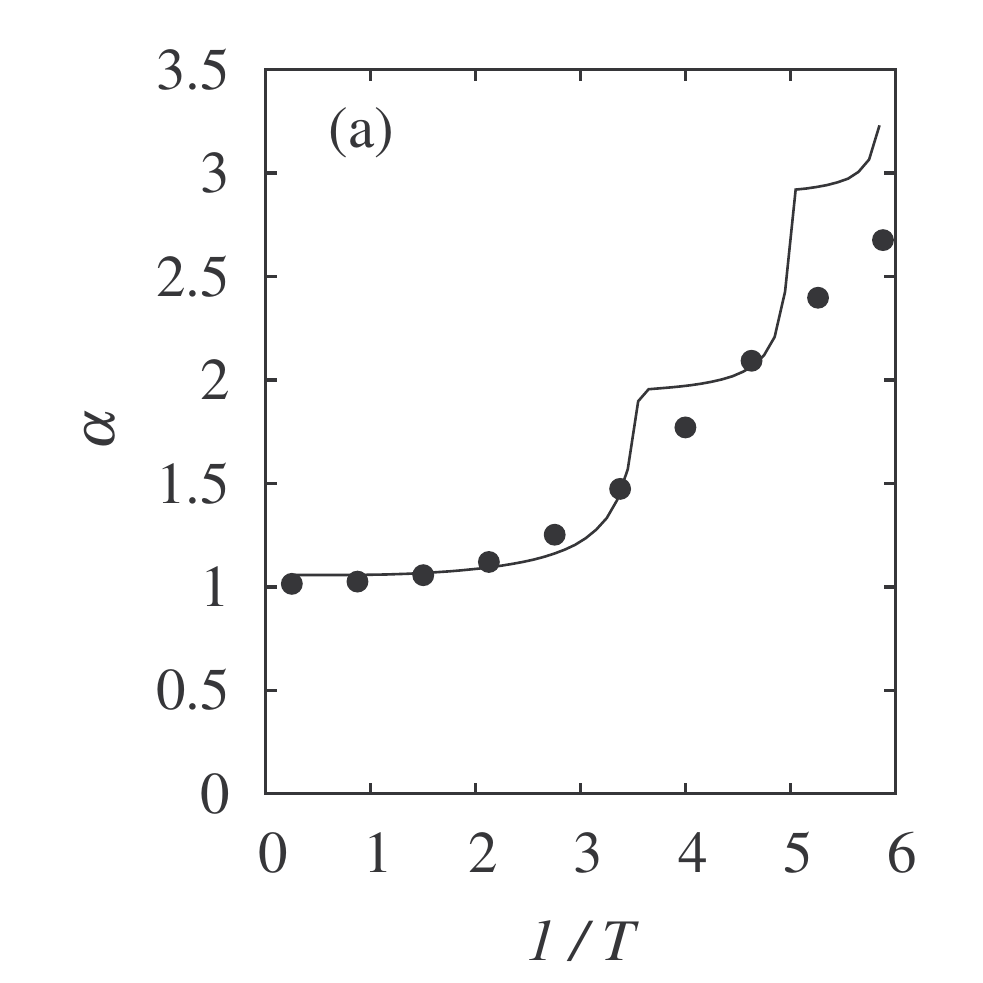}
\includegraphics[width=0.49\columnwidth]{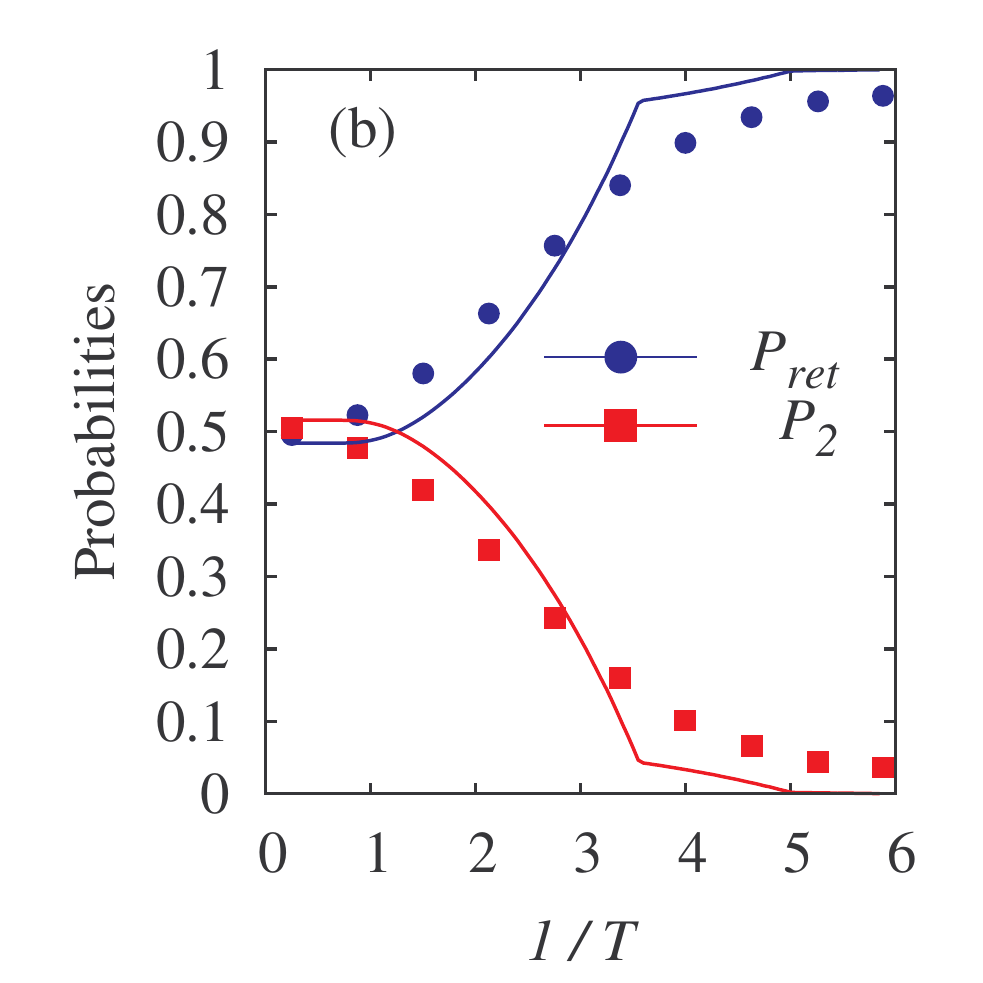}
\caption{Scaling exponent $\alpha$ (a), and probabilities of returning ($P_{ret}$) and non-returning ($P_2$) hops (b) against $1/T$ for the DPLM  from kinetic Monte Carlo simulations in Ref. \cite{lam2017dplm} (symbols) and random-tree theoretical results in \eqr{P2}{Pret} (lines). The random-tree theory uses the same parameters as in \fig{FigD} and has no additional adjustable parameter.
}
\label{Figalpha}
\label{FigPret}
\end{figure}

In \fig{Figalpha}(a), the symbols show the scaling exponent $\alpha$ defined in \eq{scaling} measured from kinetic Monte Carlo simulations for $\phiv\le 0.05$ from Ref. \cite{lam2017dplm}. We have also obtained theoretical values of $\alpha$ from the slopes of the theoretical lines of $D$ versus $\phiv$ in the log-log plot in Fig. \ref{FigD}(b). The result is plotted as a line in \fig{Figalpha}(a).
As seen, the theoretical result agrees with values from kinetic Monte Carlo simulations.

We have suggested back-and-forth particle hops at low $T$ as the main cause of kinetic arrest in glass  \cite{lam2015,lam2017dplm}. 
To study back-and-forth motions quantitatively, one can monitor the subsequent motion of a particle after it has hopped to check if it first performs a returning hop to the original site or a genuine second hop to a new site. The two types of events contribute to the probabilities $P_{ret}$ and $P_2$ respectively. 
Symbols in \fig{FigPret}(b) shows $P_{ret}$ and $P_2$ from kinetic Monte Carlo simulations of the DPLM from Ref.  \cite{lam2017dplm}. 

\newcommand{\A}{$k_0$}
We now calculate $P_{ret}$ and $P_2$  analytically. Consider a hopped particle labeled as $k_0$. Noting that a region with more voids generates proportionately more hops, particle $k_0$ resides at a region with $m$ voids with probability $p_m \propto m f(m;\barmv)$. Since Poisson distribution satisfies
\begin{equation}
\label{Poissonmean}
\sum_{m=0}^{\infty} m f(m;\barmv) = \barmv,
\end{equation}
a normalization gives
\begin{equation}
\label{pm}
p_m={m f(m;\barmv)}/{\barmv}.
\end{equation}
Denote the configuration before the first hop by the root of the random tree without loss of generality. After the first hop, the configuration is at level 1 of the tree. Returning to the root implies a return hop of particle $k_0$. Alternatively, further increments in level in the tree involve the hops of some of the $n_\Vol=\Vol/\Omega-m$ particles in the region. It may involve particle $k_0$ again after visiting on average $n_\Vol/\l$ other distinct nodes on the tree. Since $n_\Vol/\l\gg 1$, this occurs practically at an infinite tree level. For a finite tree, this is impossible and particle $k_0$ can hence only return to the root contributing to $P_{ret}$.
For an infinite tree, the configuration may first return to the root with a probability $1/c_m$ \cite{hughes1982} and this contributes to $P_{ret}$. Otherwise, it goes to a practically infinite level first with a probability $1-1/c_m$. 
The first edge encountered at a high tree-level involving particle $k_0$ pushes it to any of the $z_0$ NN sites, where $z_0$ is the particle coordination number in the physical space. Particle $k_0$ may then accidentally reverse the original hop without reversing the regional configuration with a probability $1/z_0$ and contributes to $P_{ret}$. Otherwise, it may perform a new hop with probability $1-1/z_0$ and contributes to $P_2$. Putting all these  together, particle $k_0$ first performs a new hop only for the case of an infinite tree, i.e. $c_m>1$, with a probability $(1-1/z_0) \times (1-1/c_m)$. Averaging over regions with all possible values of $m$ weighted by $p_m$ in \eq{pm}, we get
\begin{eqnarray}
\label{P2}
P_{2} = \left(1-\frac{1}{z_0}\right) \sum_{m>1/c_1} \frac{m f(m;\barmv)}{\barmv} \left( 1 - \frac{1}{mc_1} \right ), ~~~~~~
\end{eqnarray}
where \eq{cm} have been used.
Also, $P_{ret}$ follows from 
\begin{eqnarray}
\label{Pret}
P_{ret}=1-P_2.
\end{eqnarray}
Similar to $D$ in \eq{D}, the above expressions for $P_{ret}$ and $P_2$ are piecewise analytic (see \App{nonanalytic}).

We have calculated theoretical values of $P_{ret}$ and $P_2$ using \eqr{P2}{Pret} for the DPLM with $z_0=z$ and the same parameters used above including $\CDE=1.7$ and $\Vol=12$ determined in the previous fit to $D$. There is no adjustable parameter at all.
The results are shown as lines in \Fig{FigPret}(b). Reasonable quantitative agreement between \eqr{P2}{Pret}  and kinetic Monte Carlo simulations is observed. 
Since $D$ is a measure very different from $P_{ret}$ and $P_2$, without introducing any additional tunable parameter, the agreement obtained here is a highly non-trivially support of the validity of our theory for the DPLM.
Moreover,  the qualitative trend that $P_{ret}\rightarrow 1$ and $P_2 \rightarrow 0$ at low $T$ previously suggested by data from both polymer \cite{lam2015} and DPLM simulations \cite{lam2017dplm} is successfully reproduced. 

The parameters $\CDE=1.7$ and $\Vol=12$ adopted in Figs. \ref{FigD} and \ref{Figalpha} provide the best overall fit to the kinetic Monte Carlo data from Ref. \cite{lam2017dplm}. 
More generally, the calculated $D$ increases with both $\CDE$ and $\Vol$. Also, $\alpha$ decreases with increasing $\CDE$ but is less dependent on $\Vol$. 
For $D \agt 10^{-2}$ only, the best fit occurs at $\CDE=1.0$ and $\Vol=25$. 
For small $D$ at small $T$ and $\phiv$, simulated values of $D$ are in general slightly larger than those theoretically predicted. A major cause may be the stronger void aggregation tendency at low $T$ observable in DPLM simulations, which leads to deviation from \eq{Poisson} and promotes larger groups of voids and thus also the dynamics.

\section{Liquid and glassy limits}
\label{limits}
For a better intuitive understanding, we now examine asymptotic cases for which simpler analytic expressions can be derived.
At high $T$, all allowed hops are energetically possible so that $q=1$, as can be illustrated explicitly in \eq{q}.  Then, \eq{cm} implies $c_m = m(z-1) \gg 1$ for $m\ge 1$. 
Alternatively, for large $\phiv$, most regions have $m \simeq \barmv \gg 1$ voids. Then, we have $c_m=m(z-1)q \gg 1$ for not too small values of $T$ and hence $q$. 
As a result, for both cases, we have $c_m-1 \simeq c_m$. Note that the -1 term describes descending the tree level and corresponds to revisiting old configurations. Its diminished importance immediately implies the restoration of simple random walks of both the voids and the particles in the physical space.
Using \eqs{cm}{Poissonmean}, the sum in \eq{D} can be evaluated, giving a diffusion coefficient in the liquid phase as
\begin{equation}
D_{liq} \simeq \frac{z \l \w a^2    \phiv \Omega}{2d (1-\phiv\Omega)}.
\end{equation}
It follows in general a simple Arrhenius $T$ dependence at constant $\phiv$ inherent from that of $\w$. At small $\phiv$, it reduces to
\begin{equation}
D_{liq} \simeq \frac{z \l \w a^2    \phiv \Omega}{2d}.
\end{equation}
which implies the power-law in \eq{scaling} with a trivial exponent $\alpha=1$ indicating independent void motions. 
These qualitative features have been illustrated numerically in the theoretical curves for the DPLM in \fig{FigD}(a) at large $\phiv$ as well as in \fig{FigD}(b) at  high $T$.

More interestingly, we now consider the glassy limit at low $T$ and small $\phiv$. Terms in \eq{D} then have vastly different magnitudes because $D_m$ increases rapidly with $m$ while $f(m;\barmv)$ decreases rapidly. Their product maximizes sharply at a particular value of $m$, say $m^*$. 
Physically, $m^*$ is the optimal group size of coupled voids which dominates the dynamics.
A maximization of the terms in \eq{D} after applying Stirling's formula   
\begin{equation}
\label{Stirling}
m! \simeq \sqrt{2\pi} m^{m+1/2} e^{-m}
\end{equation}
gives
\begin{eqnarray}
\label{mstar1}
m^*c_1-1 = \frac {2 c_1} { \ln ( m^* / \barmv ) + 1/2m^*}.
\end{eqnarray}
Neglecting all terms except for $m=m^*$ in \eq{D}, some simple algebra using $e^{-\barmv} \simeq 1$ and Eqs. \refb{cm}, \refb{Stirling} and \refb{mstar1} gives the diffusion coefficient in the glassy limit as
\begin{eqnarray}
\label{Dglass0}
D_{glass} = 
\frac{l \w a^2 \Omega {c_1}  \left({ e \barmv/m^* }\right)^{m^*}}{2 \sqrt{2\pi} d \Vol  \sqrt{m^*} [\ln   (m^*/\barmv ) + 1/2m^*  ]} .
\end{eqnarray}
Since $c_1 \ll 1$ at low $T$, a simple approximate solution of $m^*$ from \eq{mstar1} is 
\begin{eqnarray}
\label{mstar2}
m^* \simeq 1 / {c_1}.
\end{eqnarray}
Substituting it into \eq{Dglass0} leads to
\begin{eqnarray}
\label{Dglass}
D_{glass} = 
\frac{l \w a^2 \Omega {c_1^{3/2}} \left({ e \barmv c_1  }\right)^{1/c_1}}{2 \sqrt{2\pi} d \Vol ~ [\ln   (1/\barmv c_1) + c_1/2  ]} .
\end{eqnarray}
We find numerically that $D_{glass}$ in \eq{Dglass} agrees with $D$ from \eq{D} within a factor close to order unity.

To understand how $D_{glass}$ depends on $\phiv$, we neglect weak logarithmic dependences and \eq{Dglass0} reduces to
\begin{equation}
D_{glass} \sim \phiv^{m^*}.
\end{equation}
A comparison with \eq{scaling} gives
\begin{equation}
\alpha = m^* \simeq 1/c_1
\end{equation}
where we have also used \eq{mstar2}.
Hence, the scaling exponent $\alpha$ is simply the dominant group size of the coupled voids, as expected from simple chemical kinetics. These groups of voids even at very low density are able to dominate the dynamics due to their much higher mobility. This is analogous to the role of pockets of mobile defects in the KCM's \cite{harrowell1993,garrahan2011review}.

At constant $\phiv$ and for the range of $T$ studied here, we find numerically that $D_{glass}$ exhibits a super-Arrhenius $T$ dependence which is consistent with both the Vogel-Fulcher-Tammann empirical form \cite{binderbook} and an alternative form \cite{garrahan2009,chandler2011} known to be applicable for many glassy materials.
Moreover, $D_{glass}$ admits its $T$ dependence via $\w$ and $c_1$. In particular, the factor $c_1^{1/c_1}$ is the main contributor to the super-Arrhenius slow-down. 
The precise $T$ dependence of $D_{glass}$ is relatively model dependent. For the DPLM with a uniform interaction energy distribution $g$ defined in \eq{g}, it is easy to show using \eqs{q}{c1} that $c_1 \sim q \sim T^3$ at low $T$. At constant $\phiv$ and neglecting the relatively weak non-exponential dependences, \eq{Dglass} reduces to a much simpler super-Arrhenius form
\begin{equation}
D_{glass} \sim \exp[ - (E_0 + bT^{-2})/k_BT]
\end{equation}
where $b$ is approximately a constant. One should also consider the dependence of $\phiv$ on $T$, which may be Arrhenius in the simplest cases. 
In general, other systems may admit other forms of $g$ and other $T$ dependences of $c_1$ and $\phiv$. This can generate a variety of $T$ dependences of $D_{glass}$ and may account for diverse properties such as different fragility observed in various glasses \cite{angell1995}.

\section{Conclusion} 
\label{conclusion}
In summary, we have derived a microscopic theory of glassy dynamics based on string-like hopping motions in the presence of disorder quenched in the configuration space of distinguishable particles. The elementary motions are micro-strings, which are the synchronized parts of string-like particle hopping motions. A micro-strings is enabled by a void and its propagation transports the void from one of its ends to the other. Voids can be coupled via spatially overlapping micro-strings which  interact by mutually enabling or disabling each other, noting that micro-string propagations alter the local energy landscape. The coupling can be described as a form of path interactions between voids, or equivalently a form of string interactions.
The configuration space of a local region with $m$ coupled voids is organized as a graph by identifying particle configurations as nodes and micro-string propagations as edges. The graph can be approximated by the Bethe lattice. Energetically accessible configurations constitute a random tree embedded in the Bethe lattice. 

We have analyzed the dynamics in terms of the equivalent pictures of random walks of the local configurations in the random trees in the configuration space, random walks of the voids in the physical space, and random walks of the particles in the physical space. 
At high temperature $T$, all random trees are infinite and  
all voids are fully mobile exhibiting liquid-like behaviors. As $T$ decreases, more  micro-strings become energetically unlikely. Random trees for single-void regions first go through a percolation transition and become finite implying trapping of the voids and thus also of the particles. By contrast, random trees for regions with multiple voids have more children per node  due to facilitated dynamics between voids. They can remain infinite implying mobile coupled voids. As $T$ further decreases, random trees for regions with increasingly numerous voids successively enter the non-percolating phase.
The dynamics are dominated by coupled voids of an increasing group size $m^*$. The increasing rarity of these larger dominant groups is responsible for the super-Arrhenius dynamics and kinetic arrest. The dynamics is thus characterized by a sequence of percolation transitions in the configuration space, in contrast to a single percolation transition in the physical space as one may speculate.

Explicit expressions for the particle diffusion coefficient $D$ and the particle hopping return probability $P_{ret}$ are derived. At low void density $\phiv$,  the calculated $D$ exhibits super-Arrhenius $T$-dependence typical of glasses. It also reproduces a scaling law between $D$ and $\phiv$. The scaling exponent $\alpha$ increases as $T$ decreases and is identified as the group size of the coupled voids dominating the dynamics. 
In addition, our expression for $P_{ret}$ shows a convergence  towards unity at low $T$ as observed previously in both polymer MD and DPLM simulations. 

The theory is applied quantitatively to the DPLM. Using only two tunable parameters $\CDE$ and $\Vol$, our analytical results agree well with measurements from kinetic Monte Carlo simulations of the DPLM from Ref. \cite{lam2017dplm} over a wide range of temperature and void density. 
Such a direct and detailed quantitative check of a dynamical theory with a model of glass is in our knowledge unprecedented, especially for energetically non-trivial and isotropic models in two- or three-dimensions, which have most direct physical relevance. Noting that the DPLM possesses a natural and generic definition and exhibits an extensive range of glassy properties, the agreement provides a solid support of the applicability of our theory to glasses exhibiting string-like hopping motions.

\section{Acknowledgments} We thank O.K.C. Tsui, Kai-Ming Lee, Haiyao Deng, Ho-Kei Chan, Patrick Charbonneau, Yilong Han and Giorgio Parisi for helpful discussions, and Chun-Shing Lee for technical assistance. We are grateful to  the support of Hong Kong GRF (Grant 15301014).

\appendix

\section{Non-analytic behaviors of random-tree theoretical results}
\label{nonanalytic}

Our theoretical expressions of $\tilde D$, $D$, $P_2$ and $P_{ret}$ in Eqs. \refb{Disolated}, \refb{D},  \refb{P2} and \refb{Pret} respectively are piecewise analytic functions of $T$. Isolated non-analytic points can be observed in Figs. \ref{FigD}(a), \ref{Figalpha}(a) and \ref{Figalpha}(b). 
As $T$ decreases, non-analyticities occur at $T=T_m$ at which $c_m=1$ corresponding to $c_1=1/m$ for $m=1,2,\dots$. Physically, they correspond to $T$ at which regions with $m=1, 2, \dots$ voids successively enter the immobile phase at the percolation thresholds of the random trees.

We believe that the non-analyticities represent only minor quantitative errors. They are artifacts due to two approximations.  First, although all allowed hops are in principle possible with a continuous spectrum of rates $w$ given by \eq{w}, we have considered simplified hard constraints by assuming that hops are energetically possible  with a probability $q$ defined in \eq{q} and a uniform rate $\w$ from \eq{wbar}. Revoking to the original rates $w$, all local regions with voids contribute non-vanishingly to the dynamics and all transitions must thus be corrected to crossovers. Second, we have assumed in our calculations that every region is stationary and has a constant number of voids. If one considers instead dynamic local regions and void  exchanges between local regions, the random trees admit finite life-times, which also suppress true percolation transitions. Technically, relaxing the long-time limit assumed in deriving e.g. \eq{mu} should smooth out the transitions.

\section{The approximation of independent edges in a local configuration tree}
\label{independentrate}

In \sec{mvoid}, we assume that all voids fully interact with each other so that micro-strings initiated by a void must spatially overlap with micro-strings initiated by another one (see \fig{Figint}). This is the most non-trivial assumption in our theory. We now illustrate this approximation in more detail using the example of the DPLM. In particular, we aim at showing that its impacts are quantitative rather than qualitative. A micro-string of length $l=1$ in the DPLM involves a ``hopping zone'' of $l+1$  sites comprising the original and final positions of the void. Including also the 6 nearest neighboring sites, the micro-string has an ``influence zone'' of 8 sites, corresponding to a volume $v_Z=4\Omega$ for each hopping site. The volume $\Vol$ of the local region, after including a layer of nearest neighboring  sites, expands approximately to volume $\Vol' \simeq (\Vol^{1/d} + 2\Omega^{1/d})^d$. We partition the volume $\Vol'$ into $n_Z=\Vol'/v_Z$ subregions. Neglecting spatial correlations between the hopping sites of a micro-string, we assume for simplicity that each volume $v_Z$ fits randomly into one of the $n_z$ subregions. A micro-string occupies a fraction $p_Z = (l+1)/n_Z$ of these subregions. The probability that a second micro-string of another void overlaps at least at one subregion with the first micro-string is then
\begin{equation}
  \label{Poverlap}
  P_{overlap} = 1 - ( 1 - p_Z)^{l+1} 
\end{equation}
which simplifies to
\begin{equation}
  \label{Poverlap2}
  P_{overlap} = 1 - \left( 1- \frac{4(l+1)\Omega}{\Vol'} \right)^{l+1}
\end{equation}
illustrating in particular that $P_{overlap}$ increases as $\Vol$ decreases.

Using the fitted value of $\Vol=12$ from \sec{mvoid}, we get $\Vol' \simeq 29.9$ and $P_{overlap}\simeq 0.536$. Consider for example a local region with $m=2$ voids, named A and B, with its configuration tree illustrated by \fig{Figtree2void}(b). After a hop of void A corresponding without loss of generality to descending from the root to a first-level node, there are $4m-1=7$ edges connecting this first-level node to the second level. Of these 7 edges, 3 correspond to further hops of void A and are hence new.  On average, $4P_{overlap}$ edges correspond to micro-strings of void B which overlap with and are altered by the micro-string just propagated. Thus, totally $3+4P_{overlap} \simeq 5.15$ edges
admits  new values of hopping rate $w$ (as defined in \eq{w}) which are distinct from values at the root level. 
Importantly, 5.15 new descending edges are more than only 3 edges for the single void case. This enhances the dynamics and illustrates dynamic facilitation between voids.

In contrast, on average $4(1-P_{overlap})=1.85$ edges have rates $w$ identical to those of the respective edges at the root level. They contribute to the dynamics less significantly because they do not open up possibilities of new configurations. To allow tractable calculations, we neglect that these 1.85 edges mirror those at the root level and assume all 7 edges statistically independent. This leads to the fully-interacting voids approximation.

Furthermore, the rate $w$ of none of these edges are completely independent of  each other or of those at the root level because a single hop does not completely change the energy landscape of the local region.  Neglecting also these correlations, we arrive at the independent edges approximation adopted in \sec{mvoid}. Although neglecting all these correlations is an  uncontrolled approximation, it is unlikely that it would qualitatively alter the dynamic facilitation behaviors  argued above.

The effective applicability of the fully-interacting voids approximation may be more general than that outlined above. We observe that at moderate $T$, an isolated void may be trapped and hops repeatedly around for example a dozens sites. 
 A void in a coupled pair may hop independently on average $n_I$ times before crossing the path of  the other void. All possible $n_I$-step hops of a void constitute a subgraph in the configuration graph. We envision that a subgraph may be renormalized approximately into a single edge, resulting also at the fully-interacting picture at a coarse-grained level. 

\bibliography{polymer_short}

\begin{thebibliography}{10}%
\makeatletter
\providecommand \@ifxundefined [1]{%
 \ifx #1\undefined \expandafter \@firstoftwo
 \else \expandafter \@secondoftwo
\fi
}%
\providecommand \@ifnum [1]{%
 \ifnum #1\expandafter \@firstoftwo
 \else \expandafter \@secondoftwo
\fi
}%
\providecommand \enquote [1]{``#1''}%
\providecommand \bibnamefont  [1]{#1}%
\providecommand \bibfnamefont [1]{#1}%
\providecommand \citenamefont [1]{#1}%
\providecommand\href[0]{\@sanitize\@href}%
\providecommand\@href[1]{\endgroup\@@startlink{#1}\endgroup\@@href}%
\providecommand\@@href[1]{#1\@@endlink}%
\providecommand \@sanitize [0]{\begingroup\catcode`\&12\catcode`\#12\relax}%
\@ifxundefined \pdfoutput {\@firstoftwo}{%
 \@ifnum{\z@=\pdfoutput}{\@firstoftwo}{\@secondoftwo}%
}{%
 \providecommand\@@startlink[1]{\leavevmode\special{html:<a href="#1">}}%
 \providecommand\@@endlink[0]{\special{html:</a>}}%
}{%
 \providecommand\@@startlink[1]{%
  \leavevmode
  \pdfstartlink
   attr{/Border[0 0 1 ]/H/I/C[0 1 1]}%
   user{/Subtype/Link/A<</Type/Action/S/URI/URI(#1)>>}%
  \relax
 }%
 \providecommand\@@endlink[0]{\pdfendlink}%
}%
\providecommand \url  [0]{\begingroup\@sanitize \@url }%
\providecommand \@url [1]{\endgroup\@href {#1}{\urlprefix}}%
\providecommand \urlprefix [0]{URL }%
\providecommand \Eprint[0]{\href }%
\@ifxundefined \urlstyle {%
  \providecommand \doi [1]{doi:\discretionary{}{}{}#1}%
}{%
  \providecommand \doi [0]{doi:\discretionary{}{}{}\begingroup
  \urlstyle{rm}\Url }%
}%
\providecommand \doibase [0]{http://dx.doi.org/}%
\providecommand \Doi[1]{\href{\doibase#1}}%
\providecommand \bibAnnote [3]{%
  \BibitemShut{#1}%
  \begin{quotation}\noindent
    \textsc{Key:}\ #2\\\textsc{Annotation:}\ #3%
  \end{quotation}%
}%
\providecommand \bibAnnoteFile [2]{%
  \IfFileExists{#2}{\bibAnnote {#1} {#2} {\input{#2}}}{}%
}%
\providecommand \typeout [0]{\immediate \write \m@ne }%
\providecommand \selectlanguage [0]{\@gobble}%
\providecommand \bibinfo [0]{\@secondoftwo}%
\providecommand \bibfield [0]{\@secondoftwo}%
\providecommand \translation [1]{[#1]}%
\providecommand \BibitemOpen[0]{}%
\providecommand \bibitemStop [0]{}%
\providecommand \bibitemNoStop [0]{.\EOS\space}%
\providecommand \EOS [0]{\spacefactor3000\relax}%
\providecommand \BibitemShut [1]{\csname bibitem#1\endcsname}%
\bibitem{binderbook}%
  \BibitemOpen
  \bibfield{author}{%
  \bibinfo {author} {\bibfnamefont{K.}~\bibnamefont{Binder}}\ and\ \bibinfo
  {author} {\bibfnamefont{W.}~\bibnamefont{Kob}},\ }%
  \emph{\bibinfo {title} {Glassy materials and disordered solids: An
  introduction to their statistical mechanics}}\ (\bibinfo {publisher} {World
  Scientific},\ \bibinfo {year} {2011})%
  \bibAnnoteFile{NoStop}{binderbook}%
\bibitem{ediger2012review}%
  \BibitemOpen
  \bibfield{author}{%
  \bibinfo {author} {\bibfnamefont{M.~D.}\ \bibnamefont{Ediger}}\ and\ \bibinfo
  {author} {\bibfnamefont{P.}~\bibnamefont{Harrowell}},\ }%
  \bibfield{journal}{%
  \bibinfo {journal} {J. Chem. Phys.}\ }%
  \textbf{\bibinfo {volume} {137}},\ \bibinfo {pages} {080901} (\bibinfo {year}
  {2012})%
  \bibAnnoteFile{NoStop}{ediger2012review}%
\bibitem{biroli2013review}%
  \BibitemOpen
  \bibfield{author}{%
  \bibinfo {author} {\bibfnamefont{G.}~\bibnamefont{Biroli}}\ and\ \bibinfo
  {author} {\bibfnamefont{J.~P.}\ \bibnamefont{Garrahan}},\ }%
  \bibfield{journal}{%
  \bibinfo {journal} {J. Chem. Phys.}\ }%
  \textbf{\bibinfo {volume} {138}},\ \bibinfo {pages} {12A301} (\bibinfo {year}
  {2013})%
  \bibAnnoteFile{NoStop}{biroli2013review}%
\bibitem{stillinger2013review}%
  \BibitemOpen
  \bibfield{author}{%
  \bibinfo {author} {\bibfnamefont{F.~H.}\ \bibnamefont{Stillinger}}\ and\
  \bibinfo {author} {\bibfnamefont{P.~G.}\ \bibnamefont{Debenedetti}},\ }%
  \bibfield{journal}{%
  \bibinfo {journal} {Annu. Rev. Condens. Matter Phys.}\ }%
  \textbf{\bibinfo {volume} {4}},\ \bibinfo {pages} {263} (\bibinfo {year}
  {2013})%
  \bibAnnoteFile{NoStop}{stillinger2013review}%
\bibitem{adam1965}%
  \BibitemOpen
  \bibfield{author}{%
  \bibinfo {author} {\bibfnamefont{G.}~\bibnamefont{Adam}}\ and\ \bibinfo
  {author} {\bibfnamefont{J.~H.}\ \bibnamefont{Gibbs}},\ }%
  \bibfield{journal}{%
  \bibinfo {journal} {J. Chem Phys.}\ }%
  \textbf{\bibinfo {volume} {43}},\ \bibinfo {pages} {139} (\bibinfo {year}
  {1965})%
  \bibAnnoteFile{NoStop}{adam1965}%
\bibitem{gotzebook}%
  \BibitemOpen
  \bibfield{author}{%
  \bibinfo {author} {\bibfnamefont{W.}~\bibnamefont{G{\H o}tze}},\ }%
  \emph{\bibinfo {title} {{Complex dynamics of glass-forming liquids: a
  mode-coupling theory}}}\ (\bibinfo {publisher} {Oxford University Press},\
  \bibinfo {year} {2008})%
  \bibAnnoteFile{NoStop}{gotzebook}%
\bibitem{kirkpatrick1989}%
  \BibitemOpen
  \bibfield{author}{%
  \bibinfo {author} {\bibfnamefont{T.~R.}\ \bibnamefont{Kirkpatrick}}, \bibinfo
  {author} {\bibfnamefont{D.}~\bibnamefont{Thirumalai}},\ and\ \bibinfo
  {author} {\bibfnamefont{P.~G.}\ \bibnamefont{Wolynes}},\ }%
  \bibfield{journal}{%
  \bibinfo {journal} {Phys. Rev. A}\ }%
  \textbf{\bibinfo {volume} {40}},\ \bibinfo {pages} {1045} (\bibinfo {year}
  {1989})%
  \bibAnnoteFile{NoStop}{kirkpatrick1989}%
\bibitem{fredrickson1984}%
  \BibitemOpen
  \bibfield{author}{%
  \bibinfo {author} {\bibfnamefont{G.~H.}\ \bibnamefont{Fredrickson}}\ and\
  \bibinfo {author} {\bibfnamefont{H.~C.}\ \bibnamefont{Andersen}},\ }%
  \bibfield{journal}{%
  \bibinfo {journal} {Phys. Rev. Lett.}\ }%
  \textbf{\bibinfo {volume} {53}},\ \bibinfo {pages} {1244} (\bibinfo {year}
  {1984})%
  \bibAnnoteFile{NoStop}{fredrickson1984}%
\bibitem{palmer1984}%
  \BibitemOpen
  \bibfield{author}{%
  \bibinfo {author} {\bibfnamefont{R.~G.}\ \bibnamefont{Palmer}}, \bibinfo
  {author} {\bibfnamefont{D.~L.}\ \bibnamefont{Stein}}, \bibinfo {author}
  {\bibfnamefont{E.}~\bibnamefont{Abrahams}},\ and\ \bibinfo {author}
  {\bibfnamefont{P.~W.}\ \bibnamefont{Anderson}},\ }%
  \bibfield{journal}{%
  \bibinfo {journal} {Phys. Rev. Lett.}\ }%
  \textbf{\bibinfo {volume} {53}},\ \bibinfo {pages} {958} (\bibinfo {year}
  {1984})%
  \bibAnnoteFile{NoStop}{palmer1984}%
\bibitem{ritort2003review}%
  \BibitemOpen
  \bibfield{author}{%
  \bibinfo {author} {\bibfnamefont{F.}~\bibnamefont{Ritort}}\ and\ \bibinfo
  {author} {\bibfnamefont{P.}~\bibnamefont{Sollich}},\ }%
  \bibfield{journal}{%
  \bibinfo {journal} {Adv. Phys.}\ }%
  \textbf{\bibinfo {volume} {52}},\ \bibinfo {pages} {219} (\bibinfo {year}
  {2003})%
  \bibAnnoteFile{NoStop}{ritort2003review}%
\bibitem{garrahan2011review}%
  \BibitemOpen
  \bibfield{author}{%
  \bibinfo {author} {\bibfnamefont{J.~P.}\ \bibnamefont{Garrahan}}, \bibinfo
  {author} {\bibfnamefont{P.}~\bibnamefont{Sollich}},\ and\ \bibinfo {author}
  {\bibfnamefont{C.}~\bibnamefont{Toninelli}},\ }%
  \bibinfo {journal} {in Dynamical Heterogeneities in Glasses, Colloids and
  Granular Media, edited by L. Berthier, G. Biroli, J.-P. Bouchaud, L.
  Cipelletti, and W. van Saarloosand (Oxford University Press, 2011)}%
  \bibAnnoteFile{NoStop}{garrahan2011review}%
\bibitem{glotzer1998}%
  \BibitemOpen
\bibfield{journal}{%
    }%
  \bibfield{author}{%
  \bibinfo {author} {\bibfnamefont{C.}~\bibnamefont{Donati}}, \bibinfo {author}
  {\bibfnamefont{J.~F.}\ \bibnamefont{Douglas}}, \bibinfo {author}
  {\bibfnamefont{W.}~\bibnamefont{Kob}}, \bibinfo {author}
  {\bibfnamefont{S.~J.}\ \bibnamefont{Plimpton}}, \bibinfo {author}
  {\bibfnamefont{P.~H.}\ \bibnamefont{Poole}},\ and\ \bibinfo {author}
  {\bibfnamefont{S.~C.}\ \bibnamefont{Glotzer}},\ }%
  \bibfield{journal}{%
  \bibinfo {journal} {Phys. Rev. Lett.}\ }%
  \textbf{\bibinfo {volume} {80}},\ \bibinfo {pages} {2338} (\bibinfo {year}
  {1998})%
  \bibAnnoteFile{NoStop}{glotzer1998}%
\bibitem{glotzer2003}%
  \BibitemOpen
  \bibfield{author}{%
  \bibinfo {author} {\bibfnamefont{M.}~\bibnamefont{Aichele}}, \bibinfo
  {author} {\bibfnamefont{Y.}~\bibnamefont{Gebremichael}}, \bibinfo {author}
  {\bibfnamefont{F.}~\bibnamefont{Starr}}, \bibinfo {author}
  {\bibfnamefont{J.}~\bibnamefont{Baschnagel}},\ and\ \bibinfo {author}
  {\bibfnamefont{S.}~\bibnamefont{Glotzer}},\ }%
  \bibfield{journal}{%
  \bibinfo {journal} {J. Chem Phys.}\ }%
  \textbf{\bibinfo {volume} {119}},\ \bibinfo {pages} {5290} (\bibinfo {year}
  {2003})%
  \bibAnnoteFile{NoStop}{glotzer2003}%
\bibitem{glotzer2004}%
  \BibitemOpen
  \bibfield{author}{%
  \bibinfo {author} {\bibfnamefont{Y.}~\bibnamefont{Gebremichael}}, \bibinfo
  {author} {\bibfnamefont{M.}~\bibnamefont{Vogel}},\ and\ \bibinfo {author}
  {\bibfnamefont{S.~C.}\ \bibnamefont{Glotzer}},\ }%
  \bibfield{journal}{%
  \bibinfo {journal} {J. Chem. Phys.}\ }%
  \textbf{\bibinfo {volume} {120}},\ \bibinfo {pages} {4415} (\bibinfo {year}
  {2004})%
  \bibAnnoteFile{NoStop}{glotzer2004}%
\bibitem{weeks2000}%
  \BibitemOpen
  \bibfield{author}{%
  \bibinfo {author} {\bibfnamefont{E.~R.}\ \bibnamefont{Weeks}}, \bibinfo
  {author} {\bibfnamefont{J.~C.}\ \bibnamefont{Crocker}}, \bibinfo {author}
  {\bibfnamefont{A.~C.}\ \bibnamefont{Levitt}}, \bibinfo {author}
  {\bibfnamefont{A.}~\bibnamefont{Schofield}},\ and\ \bibinfo {author}
  {\bibfnamefont{D.~A.}\ \bibnamefont{Weitz}},\ }%
  \bibfield{journal}{%
  \bibinfo {journal} {Science}\ }%
  \textbf{\bibinfo {volume} {287}},\ \bibinfo {pages} {627} (\bibinfo {year}
  {2000})%
  \bibAnnoteFile{NoStop}{weeks2000}%
\bibitem{lam2015}%
  \BibitemOpen
  \bibfield{author}{%
  \bibinfo {author} {\bibfnamefont{C.-H.}\ \bibnamefont{Lam}},\ }%
  \bibfield{journal}{%
  \bibinfo {journal} {J. Chem. Phys.}\ }%
  \textbf{\bibinfo {volume} {146}},\ \bibinfo {pages} {244906} (\bibinfo {year}
  {2017})%
  \bibAnnoteFile{NoStop}{lam2015}%
\bibitem{miyagawa1988}%
  \BibitemOpen
  \bibfield{author}{%
  \bibinfo {author} {\bibfnamefont{H.}~\bibnamefont{Miyagawa}}, \bibinfo
  {author} {\bibfnamefont{Y.}~\bibnamefont{Hiwatari}}, \bibinfo {author}
  {\bibfnamefont{B.}~\bibnamefont{Bernu}},\ and\ \bibinfo {author}
  {\bibfnamefont{J.}~\bibnamefont{Hansen}},\ }%
  \bibfield{journal}{%
  \bibinfo {journal} {J. Chem. Phys.}\ }%
  \textbf{\bibinfo {volume} {88}},\ \bibinfo {pages} {3879} (\bibinfo {year}
  {1988})%
  \bibAnnoteFile{NoStop}{miyagawa1988}%
\bibitem{vollmayr2004}%
  \BibitemOpen
  \bibfield{author}{%
  \bibinfo {author} {\bibfnamefont{K.}~\bibnamefont{Vollmayr-Lee}},\ }%
  \bibfield{journal}{%
  \bibinfo {journal} {J. Chem. Phys.}\ }%
  \textbf{\bibinfo {volume} {121}},\ \bibinfo {pages} {4781} (\bibinfo {year}
  {2004})%
  \bibAnnoteFile{NoStop}{vollmayr2004}%
\bibitem{vogel2008}%
  \BibitemOpen
  \bibfield{author}{%
  \bibinfo {author} {\bibfnamefont{M.}~\bibnamefont{Vogel}},\ }%
  \bibfield{journal}{%
  \bibinfo {journal} {Macromolecules}\ }%
  \textbf{\bibinfo {volume} {41}},\ \bibinfo {pages} {2949} (\bibinfo {year}
  {2008})%
  \bibAnnoteFile{NoStop}{vogel2008}%
\bibitem{kawasaki2013}%
  \BibitemOpen
  \bibfield{author}{%
  \bibinfo {author} {\bibfnamefont{T.}~\bibnamefont{Kawasaki}}\ and\ \bibinfo
  {author} {\bibfnamefont{A.}~\bibnamefont{Onuki}},\ }%
  \bibfield{journal}{%
  \bibinfo {journal} {Phys. Rev. E}\ }%
  \textbf{\bibinfo {volume} {87}},\ \bibinfo {pages} {012312} (\bibinfo {year}
  {2013})%
  \bibAnnoteFile{NoStop}{kawasaki2013}%
\bibitem{ahn2013}%
  \BibitemOpen
  \bibfield{author}{%
  \bibinfo {author} {\bibfnamefont{J.~W.}\ \bibnamefont{Ahn}}, \bibinfo
  {author} {\bibfnamefont{B.}~\bibnamefont{Falahee}}, \bibinfo {author}
  {\bibfnamefont{C.~D.}\ \bibnamefont{Piccolo}}, \bibinfo {author}
  {\bibfnamefont{M.}~\bibnamefont{Vogel}},\ and\ \bibinfo {author}
  {\bibfnamefont{D.}~\bibnamefont{Bingemann}},\ }%
  \bibfield{journal}{%
  \bibinfo {journal} {J. Chem. Phys.}\ }%
  \textbf{\bibinfo {volume} {138}},\ \bibinfo {pages} {12A527} (\bibinfo {year}
  {2013})%
  \bibAnnoteFile{NoStop}{ahn2013}%
\bibitem{helfferich2014}%
  \BibitemOpen
  \bibfield{author}{%
  \bibinfo {author} {\bibfnamefont{J.}~\bibnamefont{Helfferich}}, \bibinfo
  {author} {\bibfnamefont{F.}~\bibnamefont{Ziebert}}, \bibinfo {author}
  {\bibfnamefont{S.}~\bibnamefont{Frey}}, \bibinfo {author}
  {\bibfnamefont{H.}~\bibnamefont{Meyer}}, \bibinfo {author}
  {\bibfnamefont{J.}~\bibnamefont{Farago}}, \bibinfo {author}
  {\bibfnamefont{A.}~\bibnamefont{Blumen}},\ and\ \bibinfo {author}
  {\bibfnamefont{J.}~\bibnamefont{Baschnagel}},\ }%
  \bibfield{journal}{%
  \bibinfo {journal} {Phys. Rev. E}\ }%
  \textbf{\bibinfo {volume} {89}},\ \bibinfo {pages} {042603} (\bibinfo {year}
  {2014})%
  \bibAnnoteFile{NoStop}{helfferich2014}%
\bibitem{yu2017}%
  \BibitemOpen
  \bibfield{author}{%
  \bibinfo {author} {\bibfnamefont{H.-B.}\ \bibnamefont{Yu}}, \bibinfo {author}
  {\bibfnamefont{R.}~\bibnamefont{Richert}},\ and\ \bibinfo {author}
  {\bibfnamefont{K.}~\bibnamefont{Samwer}},\ }%
  \bibfield{journal}{%
  \bibinfo {journal} {Science Advances}\ }%
  \textbf{\bibinfo {volume} {3}},\ \bibinfo {pages} {e1701577} (\bibinfo {year}
  {2017})%
  \bibAnnoteFile{NoStop}{yu2017}%
\bibitem{kob1993}%
  \BibitemOpen
  \bibfield{author}{%
  \bibinfo {author} {\bibfnamefont{W.}~\bibnamefont{Kob}}\ and\ \bibinfo
  {author} {\bibfnamefont{H.~C.}\ \bibnamefont{Andersen}},\ }%
  \bibfield{journal}{%
  \bibinfo {journal} {Phys. Rev. E}\ }%
  \textbf{\bibinfo {volume} {48}},\ \bibinfo {pages} {4364} (\bibinfo {year}
  {1993})%
  \bibAnnoteFile{NoStop}{kob1993}%
\bibitem{newman1999}%
  \BibitemOpen
  \bibfield{author}{%
  \bibinfo {author} {\bibfnamefont{M.}~\bibnamefont{Newman}}\ and\ \bibinfo
  {author} {\bibfnamefont{C.}~\bibnamefont{Moore}},\ }%
  \bibfield{journal}{%
  \bibinfo {journal} {Phys. Rev. E}\ }%
  \textbf{\bibinfo {volume} {60}},\ \bibinfo {pages} {5068} (\bibinfo {year}
  {1999})%
  \bibAnnoteFile{NoStop}{newman1999}%
\bibitem{darst2010}%
  \BibitemOpen
  \bibfield{author}{%
  \bibinfo {author} {\bibfnamefont{R.~K.}\ \bibnamefont{Darst}}, \bibinfo
  {author} {\bibfnamefont{D.~R.}\ \bibnamefont{Reichman}},\ and\ \bibinfo
  {author} {\bibfnamefont{G.}~\bibnamefont{Biroli}},\ }%
  \bibfield{journal}{%
  \bibinfo {journal} {J. Chem. Phys.}\ }%
  \textbf{\bibinfo {volume} {132}},\ \bibinfo {eid} {044510} (\bibinfo {year}
  {2010})%
  \bibAnnoteFile{NoStop}{darst2010}%
\bibitem{lipson2013}%
  \BibitemOpen
  \bibfield{author}{%
  \bibinfo {author} {\bibfnamefont{N.~B.}\ \bibnamefont{Tito}}, \bibinfo
  {author} {\bibfnamefont{J.~E.}\ \bibnamefont{Lipson}},\ and\ \bibinfo
  {author} {\bibfnamefont{S.~T.}\ \bibnamefont{Milner}},\ }%
  \bibfield{journal}{%
  \bibinfo {journal} {Soft Matter}\ }%
  \textbf{\bibinfo {volume} {9}},\ \bibinfo {pages} {3173} (\bibinfo {year}
  {2013})%
  \bibAnnoteFile{NoStop}{lipson2013}%
\bibitem{lam2017dplm}%
  \BibitemOpen
  \bibfield{author}{%
  \bibinfo {author} {\bibfnamefont{L.-H.}\ \bibnamefont{Zhang}}\ and\ \bibinfo
  {author} {\bibfnamefont{C.-H.}\ \bibnamefont{Lam}},\ }%
  \bibfield{journal}{%
  \bibinfo {journal} {Phys. Rev. B}\ }%
  \textbf{\bibinfo {volume} {95}},\ \bibinfo {pages} {184202} (\bibinfo {year}
  {2017})%
  \bibAnnoteFile{NoStop}{lam2017dplm}%
\bibitem{palmer1990}%
  \BibitemOpen
  \bibfield{author}{%
  \bibinfo {author} {\bibnamefont{Ajay}}\ and\ \bibinfo {author}
  {\bibfnamefont{R.~G.}\ \bibnamefont{Palmer}},\ }%
  \bibfield{journal}{%
  \bibinfo {journal} {J. Phys. A}\ }%
  \textbf{\bibinfo {volume} {23}},\ \bibinfo {pages} {2139} (\bibinfo {year}
  {1990})%
  \bibAnnoteFile{NoStop}{palmer1990}%
\bibitem{sasa2012}%
  \BibitemOpen
  \bibfield{author}{%
  \bibinfo {author} {\bibfnamefont{S.-i.}\ \bibnamefont{Sasa}},\ }%
  \bibfield{journal}{%
  \bibinfo {journal} {Phys. Rev. Lett.}\ }%
  \textbf{\bibinfo {volume} {109}},\ \bibinfo {pages} {165702} (\bibinfo {year}
  {2012})%
  \bibAnnoteFile{NoStop}{sasa2012}%
\bibitem{rabin2016lattice}%
  \BibitemOpen
  \bibfield{author}{%
  \bibinfo {author} {\bibfnamefont{D.}~\bibnamefont{Osmanovi{\'c}}}\ and\
  \bibinfo {author} {\bibfnamefont{Y.}~\bibnamefont{Rabin}},\ }%
  \bibfield{journal}{%
  \bibinfo {journal} {J. Stat. Phys.}\ }%
  \textbf{\bibinfo {volume} {162}},\ \bibinfo {pages} {186} (\bibinfo {year}
  {2016})%
  \bibAnnoteFile{NoStop}{rabin2016lattice}%
\bibitem{rabin2015}%
  \BibitemOpen
  \bibfield{author}{%
  \bibinfo {author} {\bibfnamefont{L.~S.}\ \bibnamefont{Shagolsem}}, \bibinfo
  {author} {\bibfnamefont{D.}~\bibnamefont{Osmanovi{\'c}}}, \bibinfo {author}
  {\bibfnamefont{O.}~\bibnamefont{Peleg}},\ and\ \bibinfo {author}
  {\bibfnamefont{Y.}~\bibnamefont{Rabin}},\ }%
  \bibfield{journal}{%
  \bibinfo {journal} {J. Chem. Phys.}\ }%
  \textbf{\bibinfo {volume} {142}},\ \bibinfo {pages} {051104} (\bibinfo {year}
  {2015})%
  \bibAnnoteFile{NoStop}{rabin2015}%
\bibitem{rabin2016}%
  \BibitemOpen
  \bibfield{author}{%
  \bibinfo {author} {\bibfnamefont{L.~S.}\ \bibnamefont{Shagolsem}}\ and\
  \bibinfo {author} {\bibfnamefont{Y.}~\bibnamefont{Rabin}},\ }%
  \bibfield{journal}{%
  \bibinfo {journal} {J. Chem. Phys.}\ }%
  \textbf{\bibinfo {volume} {144}},\ \bibinfo {pages} {194504} (\bibinfo {year}
  {2016})%
  \bibAnnoteFile{NoStop}{rabin2016}%
\bibitem{tanaka2016}%
  \BibitemOpen
  \bibfield{author}{%
  \bibinfo {author} {\bibfnamefont{T.~S.}\ \bibnamefont{Ingebrigtsen}}\ and\
  \bibinfo {author} {\bibfnamefont{H.}~\bibnamefont{Tanaka}},\ }%
  \bibfield{journal}{%
  \bibinfo {journal} {J. Phys. Chem. B}\ }%
  \textbf{\bibinfo {volume} {120}},\ \bibinfo {pages} {7704} (\bibinfo {year}
  {2016})%
  \bibAnnoteFile{NoStop}{tanaka2016}%
\bibitem{douglas2013}%
  \BibitemOpen
  \bibfield{author}{%
  \bibinfo {author} {\bibfnamefont{B.~A.~P.}\ \bibnamefont{Betancourt}},
  \bibinfo {author} {\bibfnamefont{J.~F.}\ \bibnamefont{Douglas}},\ and\
  \bibinfo {author} {\bibfnamefont{F.~W.}\ \bibnamefont{Starr}},\ }%
  \bibfield{journal}{%
  \bibinfo {journal} {Soft Matter}\ }%
  \textbf{\bibinfo {volume} {9}},\ \bibinfo {pages} {241} (\bibinfo {year}
  {2013})%
  \bibAnnoteFile{NoStop}{douglas2013}%
\bibitem{swayamjyoti2014}%
  \BibitemOpen
  \bibfield{author}{%
  \bibinfo {author} {\bibfnamefont{S.}~\bibnamefont{Swayamjyoti}}, \bibinfo
  {author} {\bibfnamefont{J.}~\bibnamefont{L{\"o}ffler}},\ and\ \bibinfo
  {author} {\bibfnamefont{P.~M.}\ \bibnamefont{Derlet}},\ }%
  \bibfield{journal}{%
  \bibinfo {journal} {Phys. Rev. B}\ }%
  \textbf{\bibinfo {volume} {89}},\ \bibinfo {pages} {224201} (\bibinfo {year}
  {2014})%
  \bibAnnoteFile{NoStop}{swayamjyoti2014}%
\bibitem{chandler2011}%
  \BibitemOpen
  \bibfield{author}{%
  \bibinfo {author} {\bibfnamefont{A.~S.}\ \bibnamefont{Keys}}, \bibinfo
  {author} {\bibfnamefont{L.~O.}\ \bibnamefont{Hedges}}, \bibinfo {author}
  {\bibfnamefont{J.~P.}\ \bibnamefont{Garrahan}}, \bibinfo {author}
  {\bibfnamefont{S.~C.}\ \bibnamefont{Glotzer}},\ and\ \bibinfo {author}
  {\bibfnamefont{D.}~\bibnamefont{Chandler}},\ }%
  \bibfield{journal}{%
  \bibinfo {journal} {Phys. Rev. X}\ }%
  \textbf{\bibinfo {volume} {1}},\ \bibinfo {pages} {021013} (\bibinfo {year}
  {2011})%
  \bibAnnoteFile{NoStop}{chandler2011}%
\bibitem{cohen1961}%
  \BibitemOpen
  \bibfield{author}{%
  \bibinfo {author} {\bibfnamefont{D.}~\bibnamefont{Turnbull}}\ and\ \bibinfo
  {author} {\bibfnamefont{M.~H.}\ \bibnamefont{Cohen}},\ }%
  \bibfield{journal}{%
  \bibinfo {journal} {J. Chem. Phys.}\ }%
  \textbf{\bibinfo {volume} {34}},\ \bibinfo {pages} {120} (\bibinfo {year}
  {1961})%
  \bibAnnoteFile{NoStop}{cohen1961}%
\bibitem{han2017}%
  \BibitemOpen
  \bibfield{author}{%
  \bibinfo {author} {\bibfnamefont{X.}~\bibnamefont{Cao}}, \bibinfo {author}
  {\bibfnamefont{H.}~\bibnamefont{Zhang}},\ and\ \bibinfo {author}
  {\bibfnamefont{Y.}~\bibnamefont{Han}},\ }%
  \bibfield{journal}{%
  \bibinfo {journal} {Nat. Comm.}\ }%
  \textbf{\bibinfo {volume} {8}},\ \bibinfo {pages} {362} (\bibinfo {year}
  {2017})%
  \bibAnnoteFile{NoStop}{han2017}%
\bibitem{starr2002}%
  \BibitemOpen
  \bibfield{author}{%
  \bibinfo {author} {\bibfnamefont{F.~W.}\ \bibnamefont{Starr}}, \bibinfo
  {author} {\bibfnamefont{S.}~\bibnamefont{Sastry}}, \bibinfo {author}
  {\bibfnamefont{J.~F.}\ \bibnamefont{Douglas}},\ and\ \bibinfo {author}
  {\bibfnamefont{S.~C.}\ \bibnamefont{Glotzer}},\ }%
  \bibfield{journal}{%
  \bibinfo {journal} {Phys. Rev. Lett.}\ }%
  \textbf{\bibinfo {volume} {89}},\ \bibinfo {pages} {125501} (\bibinfo {year}
  {2002})%
  \bibAnnoteFile{NoStop}{starr2002}%
\bibitem{weitz2005}%
  \BibitemOpen
  \bibfield{author}{%
  \bibinfo {author} {\bibfnamefont{J.}~\bibnamefont{Conrad}}, \bibinfo {author}
  {\bibfnamefont{F.~W.}\ \bibnamefont{Starr}},\ and\ \bibinfo {author}
  {\bibfnamefont{D.}~\bibnamefont{Weitz}},\ }%
  \bibfield{journal}{%
  \bibinfo {journal} {J. Phys. Chem. B}\ }%
  \textbf{\bibinfo {volume} {109}},\ \bibinfo {pages} {21235} (\bibinfo {year}
  {2005})%
  \bibAnnoteFile{NoStop}{weitz2005}%
\bibitem{harrowell2006}%
  \BibitemOpen
  \bibfield{author}{%
  \bibinfo {author} {\bibfnamefont{A.}~\bibnamefont{Widmer-Cooper}}\ and\
  \bibinfo {author} {\bibfnamefont{P.}~\bibnamefont{Harrowell}},\ }%
  \bibfield{journal}{%
  \bibinfo {journal} {J. Non-Cryst. solids}\ }%
  \textbf{\bibinfo {volume} {352}},\ \bibinfo {pages} {5098} (\bibinfo {year}
  {2006})%
  \bibAnnoteFile{NoStop}{harrowell2006}%
\bibitem{newmanbook}%
  \BibitemOpen
  \bibfield{author}{%
  \bibinfo {author} {\bibfnamefont{M.}~\bibnamefont{Newman}},\ }%
  \emph{\bibinfo {title} {Networks: an introduction}}\ (\bibinfo {publisher}
  {OUP Oxford},\ \bibinfo {year} {2010})%
  \bibAnnoteFile{NoStop}{newmanbook}%
\bibitem{cassi1989}%
  \BibitemOpen
  \bibfield{author}{%
  \bibinfo {author} {\bibfnamefont{D.}~\bibnamefont{Cassi}},\ }%
  \bibfield{journal}{%
  \bibinfo {journal} {Euro. Phys. Lett.}\ }%
  \textbf{\bibinfo {volume} {9}},\ \bibinfo {pages} {627} (\bibinfo {year}
  {1989})%
  \bibAnnoteFile{NoStop}{cassi1989}%
\bibitem{anakinunpub}%
  \BibitemOpen
  \bibfield{author}{%
  \bibinfo {author} {\bibfnamefont{C.~S.}\ \bibnamefont{Lee}}\ and\ \bibinfo
  {author} {\bibfnamefont{C.-H.}\ \bibnamefont{Lam}},\ }%
  \bibinfo {journal} {unpublished~}%
  \bibAnnoteFile{NoStop}{anakinunpub}%
\bibitem{hughes1982}%
  \BibitemOpen
\bibfield{journal}{%
    }%
  \bibfield{author}{%
  \bibinfo {author} {\bibfnamefont{B.~D.}\ \bibnamefont{Hughes}}\ and\ \bibinfo
  {author} {\bibfnamefont{M.}~\bibnamefont{Sahimi}},\ }%
  \bibfield{journal}{%
  \bibinfo {journal} {J. Stat. Phys.}\ }%
  \textbf{\bibinfo {volume} {29}},\ \bibinfo {pages} {781} (\bibinfo {year}
  {1982})%
  \bibAnnoteFile{NoStop}{hughes1982}%
\bibitem{harrowell1993}%
  \BibitemOpen
  \bibfield{author}{%
  \bibinfo {author} {\bibfnamefont{P.}~\bibnamefont{Harrowell}},\ }%
  \bibfield{journal}{%
  \bibinfo {journal} {Phys. Rev. E}\ }%
  \textbf{\bibinfo {volume} {48}},\ \bibinfo {pages} {4359} (\bibinfo {year}
  {1993})%
  \bibAnnoteFile{NoStop}{harrowell1993}%
\bibitem{garrahan2009}%
  \BibitemOpen
  \bibfield{author}{%
  \bibinfo {author} {\bibfnamefont{Y.~S.}\ \bibnamefont{Elmatad}}, \bibinfo
  {author} {\bibfnamefont{D.}~\bibnamefont{Chandler}},\ and\ \bibinfo {author}
  {\bibfnamefont{J.~P.}\ \bibnamefont{Garrahan}},\ }%
  \bibfield{journal}{%
  \bibinfo {journal} {J. Phys. Chem. B}\ }%
  \textbf{\bibinfo {volume} {113}},\ \bibinfo {pages} {5563} (\bibinfo {year}
  {2009})%
  \bibAnnoteFile{NoStop}{garrahan2009}%
\bibitem{angell1995}%
  \BibitemOpen
  \bibfield{author}{%
  \bibinfo {author} {\bibfnamefont{C.~A.}\ \bibnamefont{Angell}},\ }%
  \bibfield{journal}{%
  \bibinfo {journal} {Science}\ }%
  \textbf{\bibinfo {volume} {267}},\ \bibinfo {pages} {1924} (\bibinfo {year}
  {1995})%
  \bibAnnoteFile{NoStop}{angell1995}%
\end{thebibliography}%
\end{document}